\documentclass[12pt]{article} 

\usepackage{epsfig}
\usepackage{graphicx}
\usepackage{comment}
\usepackage{latexsym}
\usepackage{hyperref}
\usepackage{amsmath}
\usepackage{color}
\usepackage{amsbsy}
\usepackage{amssymb}
\usepackage{amsthm}
\usepackage{amsfonts}
\usepackage{cite}

\newcommand{\Ms}{M_{\rm s}}

\newcommand{\m}{m_{3/2}}
\newcommand{\nF}{n_{\rm F}}
\newcommand{\nB}{n_{\rm B}}
\newcommand{\Vone}{\V_{\mbox{\scriptsize 1-loop}}}

\newcommand{\be}[0]{\begin{equation}}
\newcommand{\ee}[0]{\end{equation}}

\newcommand{\F}{{\cal F}}
\newcommand{\I}{{\cal I}}
\newcommand{\J}{{\cal J}}
\newcommand{\N}{{\cal N}}
\newcommand{\V}{{\cal V}}

\newcommand{\G}{{\cal G}}
\renewcommand{\H}{{\cal H}}

\newcommand{\Z}{\mathbb{Z}}

\newcommand{\Ka}{K{\"a}hler }
\renewcommand{\O}{{\cal O}}

\renewcommand{\Im}{{\rm Im}\,}

\newcommand{\abs}{|}
\newcommand{\Tr}{\textrm{Tr}\,}
\newcommand{\Str}{\textrm{Str}\,}

\newcommand{\ie}{{\em i.e.} }

\newcommand{\via}{{\it via} }

\newcommand{\where}{\mbox{where}}

\renewcommand{\and}{\mbox{and}}

\newcommand{\esp}{\!\!\!\phantom{\Big\abs}}
\newcommand{\desp}{\!\!\!\phantom{\Bigg\abs}}
\newcommand{\tesp}{\!\!\! \phantom{\underset{\hat \abs}{\Big\abs}}}

\newcommand{\bm}{\boldmath} 

\catcode`\@=11
\def\marginnote#1{}
\newcount\hour
\newcount\minute
\newtoks\amorpm
\hour=\time\divide\hour by60 \minute=\time{\multiply\hour by60
\global\advance\minute by-\hour}
\edef\standardtime{{\ifnum\hour<12 \global\amorpm={am}%
        \else\global\amorpm={pm}\advance\hour by-12 \fi
        \ifnum\hour=0 \hour=12 \fi
        \number\hour:\ifnum\minute<10 0\fi\number\minute\the\amorpm}}
\edef\militarytime{\number\hour:\ifnum\minute<10 0\fi\number\minute}
\def\draftlabel#1{{\@bsphack\if@filesw {\let\thepage\relax
   \xdef\@gtempa{\write\@auxout{\string
      \newlabel{#1}{{\@currentlabel}{\thepage}}}}}\@gtempa
   \if@nobreak \ifvmode\nobreak\fi\fi\fi\@esphack}
        \gdef\@eqnlabel{#1}}
\def\@eqnlabel{}
\def\@vacuum{}
\def\draftmarginnote#1{\marginpar{\raggedright\scriptsize\tt#1}}
\def\draft{\oddsidemargin -.2truein
        \def\@oddfoot{\sl preliminary draft \hfil
        \rm\thepage\hfil\sl\today\quad\militarytime}
        \let\@evenfoot\@oddfoot \overfullrule 3pt
        \let\label=\draftlabel
        \let\marginnote=\draftmarginnote
   \def\@eqnnum{(\theequation)\rlap{\kern\marginparsep\tt\@eqnlabel}%
\global\let\@eqnlabel\@vacuum}  }
\def\thebibliography#1{
\vskip 0.5cm \centerline{\bf \Large References}
\list{
[\arabic{enumi}]}{\settowidth\labelwidth{[#1]}
\leftmargin\labelwidth
\advance\leftmargin\labelsep
\usecounter{enumi}}
\def\newblock{\hskip .11em plus .33em minus .07em}
\sloppy\clubpenalty4000\widowpenalty4000
\sfcode`\.=1000\relax}

\renewcommand{\theequation}{\arabic{section}.\arabic{equation}}
\renewcommand{\section}{\setcounter{equation}{0}\@startsection
{section}{1}{0mm}{-\baselineskip}{0.5\baselineskip} {\normalfont\Large\bfseries}}
\renewcommand{\subsection}{\@startsection
{subsection}{2}{0mm}{-\baselineskip}{0.5\baselineskip} {\normalfont\large\bfseries}}
\renewcommand{\subsubsection}{\@startsection
{subsubsection}{3}{0mm}{-\baselineskip}{0.5\baselineskip}
{\normalfont\normalsize\slshape}}

\begin{document}


\begin{titlepage}
\begin{flushright}
LPTENS--16/09,
CPHT--RR080.122016
December   2016
\vspace{0.5cm}
\end{flushright}
\begin{centering}
{\bm\bf \Large $\N=2\to 0$ super no-scale models\\ 
\vspace{0.2cm}and  moduli quantum stability}

\vspace{5mm}

 {\bf Costas Kounnas$^{1}$ and Herv\'e Partouche$^2$}

 \vspace{1mm}

$^1$ Laboratoire de Physique Th\'eorique,
Ecole Normale Sup\'erieure$^\dag$,\\
24 rue Lhomond, F--75231 Paris cedex 05, France\\
{\em  Costas.Kounnas@lpt.ens.fr}

$^2$  {Centre de Physique Th\'eorique, Ecole Polytechnique, CNRS, Universit\'e Paris-Saclay\\
F--91128 Palaiseau cedex, France\\
{\em herve.partouche@polytechnique.edu}}

\end{centering}
\vspace{0.1cm}
$~$\\
\centerline{\bf\Large Abstract}\\

We consider a class of heterotic $\N=2\to 0$ super no-scale  $\Z_2$-orbifold models.  An appropriate  stringy Scherk-Schwarz supersymmetry breaking induces tree level masses to all massless bosons of the twisted hypermultiplets and therefore stabilizes all twisted moduli. At high supersymmetry breaking scale, the tachyons that occur in the $\N=4\to0$ parent theories are projected out, and no Hagedorn-like instability takes place in the $\N=2\to 0$ models (for small enough marginal deformations). At low supersymmetry breaking scale, the stability of the untwisted moduli is studied at the quantum level by taking into account both untwisted and twisted contributions to the 1-loop effective potential. The latter depends on the specific branch of the gauge theory along which the background can be deformed. We derive its expression in terms of all classical marginal deformations in the pure Coulomb phase, and in some mixed Coulomb/Higgs phases. In this class of  models, the super no-scale condition requires having at the massless level equal numbers of untwisted bosonic and twisted fermionic degrees of freedom. 
Finally, we show that $\N=1\to 0$ super no-scale models are obtained by implementing a second $\Z_2$ orbifold twist on $\N=2\to 0$ super no-scale $\Z_2$-orbifold models.
\begin{quote}

\end{quote}

\vspace{3pt} \vfill \hrule width 6.7cm \vskip.1mm{\small \small \small
  \noindent
   $^\dag$\ Unit{\'e} mixte  du CNRS et de l'Ecole Normale Sup{\'e}rieure associ\'ee \`a l'Universit\'e Pierre et Marie Curie (Paris 6), UMR 8549.}\\

\end{titlepage}
\newpage
\setcounter{footnote}{0}
\renewcommand{\thefootnote}{\arabic{footnote}}
 \setlength{\baselineskip}{.7cm} \setlength{\parskip}{.2cm}

\setcounter{section}{0}


\section{Introduction}

In string theory, even when starting classically in a flat four-dimensional background, the vacuum energy induced at the quantum level is hard to reconcile with the present cosmological constant. When supersymmetry is hardly broken, the 1-loop effective potential is generically of order $\Ms^4$, where $\Ms$ is the string scale, which is far too large. On the contrary, if supersymmetry is exact, the quantum potential vanishes identically at least at the perturbative level, or leads non-perturbatively to an anti de Sitter vacuum with restored supersymmetry.  {\it A priori} more promising, the no-scale models \cite{noscale} consist somehow of an intermediate situation. At the classical level, these backgrounds realize in flat space a spontaneous breaking of  supersymmetry at a scale $\m$, which is a flat direction of the tree level potential. However, if the order of magnitude of the quantum effective potential is dictated by $\m$, it happens to be generically too large. Moreover, the quantum potential induces tadpoles for the classical moduli fields, including the dilaton and the ``no-scale modulus" parameterized  by $\m$, which are responsible for a destabilization of the flat background. 

Some exceptions however exist, at least at the 1-loop level, when the spontaneous breaking of supersymmetry arises \via  ``coordinate dependent compactification" \cite{SSstring, Kounnas-Rostand}, the stringy version of the Scherk-Schwarz mechanism \cite{SS}. This total breaking of supersymmetry can be  implemented on initially $\N=4$, 2 or 1 heterotic or type II orbifold models, as well as on orientifold theories \cite{openSS}, or on marginally deformed fermionic constructions \cite{Kounnas:1990ww, N=0thresh}. In this framework, some theories referred as super no-scale models \cite{ADM,planck2015,SNSM,FR} induce an exponentially suppressed 1-loop vacuum energy, whose order of magnitude can  easily be of order (or lower than) the presently observed cosmological constant. Type II \cite{L=0} and orientifold \cite{1-L=0} theories with exactly vanishing  vacuum energy at 1-loop even exist.  In all known examples, the models arise at extrema of the quantum effective potential, with respect to all directions that are lifted. However, the question of the stability of the non-flat directions must be addressed. In other words, does the model sit at a minimum, maximum or saddle point of its potential~? This problem has been addressed  for the super no-scale models realizing the $\N=4\to 0$ spontaneous breaking of supersymmetry \cite{SNSM}  and is reconsidered for less symmetric theories in the present work, such as models implementing an $\N=2\to 0$ breaking. 

In Ref. \cite{SNSM}, one considers the $\N=4\to 0$ no-scale models for given internal metric, antisymmetric tensor and Wilson lines background expectation values. Supposing the point in moduli space is such that no mass scale below $\m$ exist, we denote $c\Ms$ the lowest mass scale above $\m$. In this case, the 1-loop effective potential  takes the form \cite{AntoniadisTeV}
\be
\label{V}
\Vone^{\N=4\to 0}=\xi (\nF-\nB)\,  \m^4+\O\!\left(c^2\Ms^2\m^2\, e^{-c\Ms/\m}\right)\!,
\ee
where the gravitino mass $\m$ scales inversely to the volume involved in the stringy Scherk-Schwarz mechanism and $\nF, \nB$ are the numbers of massless fermionic  and bosonic degrees of freedom. The $\m^4$ dominant contribution arises from the light towers of Kaluza-Klein states, whose masses are of order $\m$, while $\xi>0$ depends on moduli other than the dilaton and $\m$. For the quantum vacuum energy and tadpoles to be exponentially small, one can focus on the models satisfying the super no-scale condition $\nF=\nB$ \cite{ADM,planck2015,SNSM}. In this case, the 1-loop vacuum energy is of the order of the observed cosmological constant, provided the gravitino mass $\m$ is about 2 orders of magnitude smaller than the scale $c\Ms$. However, switching on small marginal deformations, collectively denoted $Y$, around the point in classical moduli space we started with, one induces new mass scales lower than $\m$. Some of the $\nF+\nB$ initially massless states  acquire small masses. When the mass scales $Y\!\Ms$ reach the order of~$\m$, the exponential contributions in Eq. (\ref{V}) are $\O(\m^4)$, thus correcting $\nF, \nB$ which now take new integer values. In other words, $\nF, \nB$ are effectively functions of $Y$ which interpolate between different integer values corresponding to distinct massless spectra.

 To study the local stability of an  $\N=4\to 0$ super no-scale model \cite{SNSM} around a point in moduli space characterized by  integer $\nF$ and $\nB$, one has to expand these two functions at quadratic order in~$Y$. Due to the underlying $\N=4$ structure, the moduli deformations $Y$ are Wilson lines along~$T^6$. The result is that those which are associated to non-asymptotically free gauge group factors become tachyonic at 1-loop. They condense, break spontaneously the associated gauge symmetry which enters a Coulomb branch, and induce a destabilization of the vacuum. On the contrary, the Wilson line  associated to asymptotically free gauge group factors become massive and are dynamically  attracted to $Y=0$. The Wilson lines associated to conformal groups remain massless.  

In Sect. \ref{2->0}, we consider $\N=2\to 0$ super no-scale theories realized as $\Z_2$-orbifolds of $\N=4\to 0$ no-scale models. At the level of exact $\N=2$ supersymmetry, the twisted hypermultiplets introduce new moduli fields living on a quaternionic manifold. We show that the implementation of the stringy Scherk-Schwarz mechanism can  always be chosen so that all twisted moduli acquire a tree level mass of order $\m$ and are no more marginal in the non-supersymmetric theory. In these models, the super no-scale condition amounts to having classically equal numbers of massless untwisted bosonic and twisted fermionic degrees of freedom. Moreover, the tachyons that appear at tree level in the parent $\N=4\to 0$ theory \cite{Kounnas-Rostand} when  $\m$ is of the order of $\Ms$ are automatically projected out in the $\N=2\to 0$ models. In other words, when $\m$ is large and local perturbations of other moduli are allowed, no Hagedorn-like instability occurs.  

In Sect.~\ref{defor}, we  study the local stability of the untwisted  moduli in this class of super no-scale models. The analysis  generalizes that of Ref. \cite{SNSM} by taking into account, in the 1-loop effective potential~$\Vone^{\N=2\to 0}$, the contributions arising from the twisted fermions. The expression of $\Vone^{\N=2\to 0}$, which we determine at second order in moduli fields, is distinct in each branch of the gauge theory along which the classical background can be deformed. To be more specific, we derive the quantum potential as a function of all moduli fields in the pure Coulomb branch, as well as in some mixed Coulomb/Higgs branches. Moreover, we show that in this class of models, because all moduli fields are untwisted, the number of marginal deformations in any branch of the gauge theory is universal once the model is compactified down to two dimensions. 
  
In Sect.~\ref{1->0}, we show  that $\N=2\to 0$ super no-scale $\Z_2$-orbifold models can automatically lead descendent $\N=1\to 0$ super no-scale theories, by implementing a second $\Z_2$ orbifold twist. However, we argue that the analysis of the background stability must be generalized to include new twisted moduli deformations. 

A summary of our hypothesis and results can be found in the conclusion, Sect.~\ref{cl}. 


\section{\bm A class of $\N=2\to 0$ super no-scale models}
\label{2->0}

In this section, we construct $\N=2\to 0$  super no-scale backgrounds, keeping in mind the goal of Sect. \ref{defor}, which is to study their stability at the quantum level. In the framework of heterotic $\Z_2$-orbifold compactifications, at the exact $\N=2$ level, the models admit special \Ka moduli belonging to vector multiplets arising in the untwisted sector. Quaternionic deformations belonging to hypermultiplets also exist and occur generically in both untwisted and twisted sectors. In the following, we highlight a particular class of models characterised by an implementation of the stringy Scherk-Schwarz  supersymmetry breaking that lifts classically all moduli of the twisted sector. These models are generic in the sense that both types of moduli, special \Ka and quaternionic, are allowed. However, they are also particular, since the complex structure of the internal space cannot be deformed away from the orbifold point, as follows from the non-existence of twisted deformations \cite{KV}. 

We consider heterotic no-scale models on $T^2\times T^4/\Z_2$, where the $\N=2\to 0$ spontaneous breaking of supersymmetry is implemented by a coordinate dependent compactification on~$T^2$. We denote the spacetime, $T^2$ and $T^4$ coordinates as $X^{0,1,2,3}$, $X^{4,5}$ and $X^{6,7,8,9}$, respectively. For notational convenience, we restrict to the case where the stringy Scherk-Schwarz mechanism is implemented along the compact direction $X^4$ only, which is supposed to be large, for $\m$ to be lower than $\Ms$. Moreover, even if it is not necessary, we will quote our results in the case where the second direction of $T^2$ is also large. Our aim is twofold : 

$\bullet$ First, we want models that develop a super no-scale structure. In terms of the $\mbox{1-loop}$ partition function $Z$, the effective potential can be expressed as an integral over the fundamental domain $\F$ of $SL(2,\Z)$,  
\be
\label{vi}
\V_{\mbox{\scriptsize 1-loop}}^{\N=2\to 0}= -{M_{\rm s}^4\over (2\pi)^4}\int_\F {d^2\tau\over 2\tau_2^2}\, Z\, ,
\ee 
where $\tau= \tau_1+i\tau_2$ is the Techm\"uller parameter. As long as a model sits at a point in moduli space where no mass scale is lower than $\m$, the untwisted and twisted sectors both yield  contributions as shown in Eq. (\ref{V}), so that 
\be
\V_{\mbox{\scriptsize 1-loop}}^{\N=2\to 0}= \xi (\nF^{\rm u}+\nF^{\rm t}-\nB^{\rm u}-\nB^{\rm t})\,  \m^4+\O\!\left(c^2\Ms^2\m^2\, e^{-c\Ms/\m}\right)\!.
\ee 
In the above expression, $\nF^{\rm u},\nB^{\rm u}$ are the numbers of massless untwisted fermions and bosons, while $\nF^{\rm t},\nB^{\rm t}$ are their counterparts in the twisted sector. For a model to be super no-scale, we  require 
\be
\label{snc}
\nF^{\rm u}+\nF^{\rm t}=\nB^{\rm u}+\nB^{\rm t}\, .
\ee

$\bullet$ Second, we want the precise implementation of the coordinate dependent compactification to imply the twisted moduli present at the $\N=2$ level to be lifted at tree level. 

Note that in the present work, the $\Z_2$ twist is {\em non-freely} acting and gives {\em a priori} rise to massless states in the twisted sector. This situation is to be contrasted with the simpler one, already studied in Ref. \cite{SNSM}, where the $\Z_2$ twist on $T^4$ also shifts the direction $X^5$ of $T^2$. In this case, there are no fixed points and the twisted states are automatically super massive (the strings are stretched along $X^5$), even at the $\N=2$ level. 


\vspace{.2cm}
{\large \em \noindent A representative model}

\noindent The starting point  to construct the simplest model that realizes the above goal is the $E_8\times E_8$ heterotic string compactified on $T^2\times T^4$. The stringy Scherk-Schwarz mechanism can be introduced by implementing a $\Z_2$ orbifold shift along $X^4$, while  a $\Z_2$ orbifold twist acts on~$X^{6,7,8,9}$. The 1-loop partition fonction is
\begin{align}
\label{Z}
Z&= {1\over 2}\sum_{H,G}Z\big[{}^H_G\big]\nonumber \\
  &={1\over 2}\sum_{H,G}\;   {1\over 2}\sum_{h,g}\;  {1\over 2}\sum_{a,b}\; S\big[{}^{a;h}_{b\, ;g}\big]\; S'\big[{}^{h;H}_{g \, ;G}\big]\;Z_{4,0}^{({\rm F})}\!\big[{}^{a;H}_{b\, ;G}\big]\; O_{2,2}^{({\rm n.c.})}\; {\Gamma_{2,2}\big[{}^{h}_{g}\big]\over \eta^2\bar \eta^2}\; Z_{4,4}\big[{}^H_G\big]\; \bar Z_{0,8}\big[{}^H_G\big]\; \bar O_{0,8}^{(E_8)},
\end{align}   
where $Z_{4,0}^{(\rm F)}\!\big[{}^{a}_{b}  \big]$ and $O_{2,2}^{({\rm n.c.})}$ are the conformal blocks arising from the left-moving worldsheet fermions and non-compact spacetime coordinates in light cone gauge, while  $\bar Z_{0,8}\big[{}^H_G\big]\bar O_{0,8}^{(E_8)}$ is the contribution of the 16 additional right-moving bosonic degrees of freedom,
\begin{align}
&Z_{4,0}^{(\rm F)}\!\big[{}^{a; H}_{b\, ;G}  \big]\!=(-1)^{a+b+ab}\, {\theta\big[{}^a_b\big]^2\over \eta^2}\, {\theta\big[{}^{a+H}_{b+G}\big]\over \eta}\, {\theta\big[{}^{a-H}_{b-G}\big]\over \eta} \,,   &&O_{2,2}^{({\rm n.c.})} ={1\over \tau_2\eta^2 \bar \eta^2}\, ,\nonumber\tesp \\
&\bar Z_{0,8}\big[{}^H_G\big]\!={1\over 2}\sum_{\gamma,\delta} {\bar \theta\big[{}^\gamma_\delta\big]^6\over \bar \eta^6}\,{\bar \theta\big[{}^{\gamma+H}_{\delta\, +G}\big]\over \bar \eta}\, {\bar \theta\big[{}^{\gamma-H}_{\delta\, -G}\big]\over \bar \eta} \, , &&\bar O_{0,8}^{(E_8)}= {1\over 2}\sum_{\gamma',\delta'} {\bar \theta\big[{}^{\gamma'}_{\delta'}\big]^8\over \bar \eta^8}\, .
\end{align}
In our conventions, the spin structures $a,b$, the twists $H,G$ and $\gamma,\delta,\gamma',\delta'$ are integer modulo~2, while our definitions of the Dedekind $\eta$ and Jocobi $\theta\big[{}^\alpha_\beta\big]$ functions can be found  in Ref.  \cite{KiritsisBook}. The conformal block associated to the $T^4/\Z_2$ directions is 
\be
\label{z44}
Z_{4,4}\big[{}^H_G\big]\!= \left\{
\begin{array}{ll}
\displaystyle (1+\Gamma_{4,4}^++\Gamma_{4,4}^-)\, {1\over \eta^4\bar\eta^4}&\mbox{if $(H,G)\equiv (0,0)\, ,$}\tesp\\
(\displaystyle 1+\Gamma_{4,4}^+-\Gamma_{4,4}^-)\, {16\eta^2\bar\eta^2 \over \theta\big[{}^1_0\big]^2\, \bar\theta\big[{}^1_0\big]^2}&\mbox{if $(H,G)\equiv (0,1)\, ,$}\tesp \\
\displaystyle {16\eta^2\bar\eta^2 \over \theta\big[{}^{\;\;\, 0}_{1-G}\big]^2\, \bar\theta\big[{}^{\;\;\, 0}_{1-G}\big]^2}&\mbox{if $H\equiv 1\, ,$}
\end{array}
\right.
\ee 
where $\Gamma_{4,4}^+$ and $\Gamma_{4,4}^-$ are the contributions of the 4-torus zero modes that are even or odd under the $\Z_2$ twist. As explained in the Appendix, they satisfy  $\Gamma_{4,4}^+=\Gamma_{4,4}^-$. Finally, the $T^2$ coordinates contribution involves  the shifted lattice
\be
 \label{lat}
\Gamma_{2,2}\big[{}^{h}_{g} \big]\!(T_1,U_1) =\!\!\!\sum_{\scriptsize\begin{array}{c}m_4,m_5\\ n_4,n_5\end{array}}\!\!\!(-1)^{gm_4}\, q^{{1\over 2}\abs p_L\abs^2}\bar q^{{1\over 2}\abs p_R\abs^2} ,
\ee
where $h,g$ are integer modulo 2, $q=e^{2i\pi\tau}$ and
\begin{align}
\label{pLR}
p_L&={1\over \sqrt{2\, \Im T_1\, \Im U_1}}\left[U_1m_4-m_5+T_1 \big(n_4+{1\over 2}h\big)+T_1 U_1 n_5\right],\desp \nonumber\\
p_R&={1\over \sqrt{2\, \Im T_1\, \Im U_1 }}\left[ U_1m_4-m_5+\bar T_1\big(n_4+{1\over 2}h\big)+\bar T_1 U_1 n_5\right] ,
\end{align}
in terms of integer momenta $m_4,m_5$ and winding numbers $n_4,n_5$, as well as the internal metric and antisymmetric tensor, through the \Ka and complex structure moduli 
\be
\label{TU}
T_1=i\sqrt{G_{44}G_{55}-G_{45}^2}+B_{54} \, , \qquad U_1={i\sqrt{G_{44}G_{55}-G_{45}^2}+G_{54}\over G_{44}}\, .
\ee

The $\N=2\to 0$ spontaneous breaking is implemented by coupling the lattice shift $h,g$ to the spin structure $a,b$, where $a=0$ ($a=1$) corresponds to spacetime bosons (fermions). This is done by inserting in the partition function the modular invariant sign \cite{Kounnas-Rostand}
\be
\label{S}
S\big[{}^{a;h}_{b\, ;g}\big]\; =(-1)^{ga+hb+gh}\, .
\ee 
Note that the light spectrum must have vanishing winding numbers along the large direction~$X^4$, which implies $h=0$ (see Eq. (\ref{pLR})) and $S=(-1)^{ga}$. If nothing else is introduced in the partition function, the initially massless bosons ($a=0$) don't see the breaking, while the massless fermions ($a=1$) acquire a mass of order $\m$. In this case, the number of massless fermions is always vanishing and the model has no chance to be super no-scale. To remedy this fact, we insert in $Z$ another modular invariant sign \cite{critical}
\be
S'\big[{}^{h;H}_{g \, ;G}\big]=(-1)^{gH+hG}\, ,
\ee 
which for $h=0$  yields $SS'=(-1)^{g(a+H)}$, so that :

$\bullet$ In the untwisted sector, $H=0$, the situation is as before.  The $\N=2\to 0$ breaking induces a tree level mass $\m$ to the massless fermions, while the massless bosons are not modified.

$\bullet$ In the twisted sector however, $H=1$, the situation is reversed. The $\N=2\to 0$ breaking induces a tree level mass $\m$ to the massless bosons, while the massless fermions are not modified.
\\
Therefore, we have 
\be
\nF^{\rm u}=0\, , \qquad \nB^{\rm t}=0\, , 
\ee 
and the super no-scale condition (\ref{snc}) becomes
\be
 \nF^{\rm t}=\nB^{\rm u}\, .
\ee 
In other words, in the 1-loop effective potential, we want the contribution of the untwisted massless sector, which is purely bosonic, to compensate that of the twisted massless sector, which is purely fermionic. 
Note that the consequences of the $SS'$ insertion in a partition function extend far beyond the particular example we consider. They are valid in any heterotic $\Z_2$-orbifold model, where the stringy Scherk-Schwarz mechanism is implemented along the untwisted directions.  

\vspace{.2cm}
{\large \em \noindent The spectrum}

\noindent In order to see how things  work in detail, we write the partition function (\ref{Z}) in terms of~$SO(2N)$ affine characters 
\begin{align}
O_{2N}&={\theta\big[{}^0_0\big]^N+\theta\big[{}^0_1\big]^N\over 2\eta^N}\, , &V_{2N}&={\theta\big[{}^0_0\big]^N-\theta\big[{}^0_1\big]^N\over 2\eta^N}\, ,\nonumber \tesp\\
S_{2N}&={\theta\big[{}^1_0\big]^N+(-i)^N\theta\big[{}^1_1\big]^N\over 2\eta^N}\, ,
&C_{2N}&={\theta\big[{}^1_0\big]^N-(-i)^N\theta\big[{}^1_1\big]^N\over 2\eta^N}\, .
\label{charac}
\end{align}
In the untwisted sector, $H=0$, we find
\begin{align}
\label{Z0}
Z\big[{}^{\, 0}_G\big]\!=& \; O_{2,2}^{({\rm n.c.})}\;Z_{4,4}\big[{}^{\, 0}_G\big]\; \Big( \bar O_{12}\bar O_{4}+\bar V_{12}(-1)^G\bar V_{4}+\bar C_{12}(-1)^G\bar C_{4}+\bar S_{12}\bar S_{4}\Big)\,  \bar O_{0,8}^{(E_8)}  \nonumber\esp  \\
&\times \Big\{ \; \; \; O_{2,2}\big[{}^0_0\big] \Big(O_4(-1)^GV_4+V_4O_4\Big)-O_{2,2}\big[{}^0_1\big] \Big(C_4(-1)^GC_4+S_4S_4\Big)\nonumber\desp  \\
& \;\;\;\;\;\,- O_{2,2}\big[{}^1_0\big] \Big(C_4(-1)^GS_4+S_4C_4\Big)+O_{2,2}\big[{}^1_1\big] \Big(O_4(-1)^GO_4+V_4V_4\Big)\; \Big\},
\end{align}
where we have defined characters associated to the shifted $T^2$ as 
 \be
 \label{defO}
O_{2,2}\big[{}^h_g\big]={\Gamma_{2,2}\big[{}^{h}_{0} \big]+(-1)^g\, \Gamma_{2,2}\big[{}^{h}_{1} \big]\over 2\eta^2\bar\eta^2}={1\over \eta^2\bar \eta^2}\!\!\!\sum_{\scriptsize\begin{array}{c}k_4,m_5\\ n_4,n_5\end{array}}\!\!\!q^{{1\over 2}\abs p_L\abs^2}\, \bar q^{{1\over 2}\abs p_R\abs^2} ,
\ee
with  momentum $m_4$ redefined as $2k_4+g$ in the expressions of $p_L,p_R$ in Eq. (\ref{pLR}). 
Similarly, one obtains in the twisted sector, $H=1$, 
\begin{align}
\label{Z1}
Z\big[{}^{\, 1}_G\big]\!=& \; O_{2,2}^{({\rm n.c.})}\;Z_{4,4}\big[{}^{\, 1}_G\big]\; \Big( \bar O_{12}(-1)^G\bar C_{4}+\bar V_{12}\bar S_{4}+\bar C_{12}\bar O_{4}+\bar S_{12}(-1)^G\bar V_{4}\Big)\,  \bar O_{0,8}^{(E_8)} \nonumber\esp  \\
& \times \Big\{ -O_{2,2}\big[{}^0_0\big] \Big(C_4O_4+S_4(-1)^GV_4\Big)+O_{2,2}\big[{}^0_1\big] \Big(O_4S_4+V_4(-1)^GC_4\Big)\nonumber\desp  \\
& \; \; \; \;\;\;\; + O_{2,2}\big[{}^1_0\big] \Big(O_4C_4+V_4(-1)^GS_4\Big)-O_{2,2}\big[{}^1_1\big] \Big(C_4V_4+S_4(-1)^GO_4\Big)\; \Big\}.
\end{align}
Some remarks are in order : 

$\bullet$  Due to the large volume of $T^2$, the sector $h=1$, which yields non-vanishing winding number $n_4+{1\over 2}$, leads contributions of order $e^{-\pi\tau_2\Im T_1/4\Im U_1}\ll 1$ for $\tau\in\F$. Thus,  all conformal blocks proportional to $O_{2,2}\big[{}^1_0\big]$ and $O_{2,2}\big[{}^1_1\big]$ will not be considered explicitly from now on. 

$\bullet$ The massless spectrum arises from the conformal blocks proportional to $O_{2,2}\big[{}^0_0\big]$, for vanishing momenta and winding numbers along $T^2$. In the untwisted (twisted) sector, as announced before, it is bosonic (fermionic). It is accompanied by towers of light bosonic (fermionic) Kaluza-Klein states, with momenta $2k_4$ and $m_5$. 

$\bullet$ The remaining light spectrum arises from the conformal blocks proportional to $O_{2,2}\big[{}^0_1\big]$. It is composed of   towers of fermionic (bosonic) Kaluza-Klein states, with momenta $2k_4+1$ and $m_5$, which are superpatners of the above mentioned states, with mass degeneration lifted. The mass gap in these sectors is the gravitino mass $\m$, which satisfies 
\begin{equation}
\label{m32}
m^2_{3/2}={\abs U_1\abs^2 M^2_{\rm s}\over \Im T_1\, \Im U_1}\, .
\end{equation} 
It vanishes in the large $T^2$ volume limit, $\Im T_1\to +\infty$, $U_1$ finite, where supersymmetry is recovered.

\vspace{.2cm}
{\large \em \noindent The untwisted sector}

\noindent In order to realize a super no-scale model, we first count the massless states in the untwisted sector, $H=0$. In a theory where the  breaking of supersymmetry is spontaneous, there cannot be any physical tachyon when the order of magnitude of $\m$ is lower than $\Ms$. We thus have
\be
{1\over 2}\left(Z\big[{}^{0}_0\big]\!+Z\big[{}^{0}_1\big]\right)\!\!\Big\abs_{\rm level \, matched}=(\nB^{\rm u}-0) (q\bar q)^0+\cdots\, ,
\ee
where the ellipsis account for the contributions of $\nB^{\rm u}$ fermionic superpartners of mass $\m$ and all more massive states. However, only $Z\big[{}^{0}_0\big]$ needs to be expanded, since the sector $h=0$ in $Z\big[{}^{0}_1\big]$ vanishes, as can be seen in  the second line of Eq. (\ref{Z0}). Not that this fact is not specific to the present model. It arises from the supersymmetry breaking sign $S$ and the identity
\be
\label{id}
{1\over 2}\sum_{a,b}S\big[{}^{a;0}_{b\, ;g}\big] (-1)^{a+b+ab}\, \theta\big[{}^a_b\big]^2\, \theta\big[{}^{a+0}_{b\, +1}\big]\, \theta\big[{}^{a-0}_{b\, -1}\big]\!=- \theta\big[{}^{\;\;1}_{1-g}\big]^2\, \theta\big[{}^{1}_{g}\big]^2=0\, .
\ee
Since $Z\big[{}^{0}_0\big]$ is the partition function of the parent $\N=4\to 0$ no-scale model, it  is more naturally expressed in terms of $SO(8)$ and $SO(16)$ affine characters, using
\be
V_8=O_4V_4+V_4O_4\, , \quad \bar O_{16}=\bar O_{12}\bar O_{4}+\bar V_{12}\bar V_{4}\, , \quad \bar S_{16}=\bar C_{12}\bar C_{4}+\bar S_{12}\bar S_{4}\, .
\ee
Defining $G^{(T^4)}$ the gauge symmetry group arising from the $T^4$ lattice on the right-moving side of the string, and reminding that $\bar O_{16}+\bar S_{16}=\bar O_{0,8}^{(E_8)}$, we find 
\be
\label{00}
{1\over 2}\, Z\big[{}^{0}_0\big]\!={1\over 2}\, {8\over \tau_2} \left({1\over\bar q}+2+2+\dim G^{(T^4)}+\dim (E_8\times E_8)+\O(q)\right) .
\ee
The first 2 in the parentheses comes from $O_{2,2}^{({\rm n.c.})}$ and account for the bosonic part of the $\N=4$ supergravity multiplet, while the second 2 is the dimension of the $U(1)^2$ right-moving gauge symmetry arising from $O_{2,2}\big[{}^0_0\big]$. We thus have 
\be
\label{nb}
\nB^{\rm u}=4\big(\!\dim G^{(T^4)}+500\big) .
\ee

In order to find the representations in which the $\nB^{\rm u}$ untwisted massless states are organized, we expand 
\be
\label{z4}
Z_{4,4}\big[{}^{\, 0}_G\big]\!={
1+\big[N^++(-1)^G(N^-\!+4)\big]\bar q+\cdots \over q^{4/24}\bar q^{4/24}}\quad \where\quad N^\pm={\dim G^{(T^4)}-4\over 2}.
\ee
$N^\pm$ are non-vanishing if the $\Gamma_{4,4}^{\pm}$ lattices moduli sit at enhanced symmetry points, while 4 is the rank of $G^{(T^4)}$.  In these notations, we find
\begin{align}
\label{part}
{1\over 2}\left(Z\big[{}^{0}_0\big]\!+ Z\big[{}^{0}_1\big]\right)\!=&\;O_{2,2}^{({\rm n.c.})}\;O_{2,2}\big[{}^0_0\big]\; {1\over q^{4/24}\bar q^{4/24}}\; \bar O_{0,8}^{(E_8)} \nonumber \\
&\times\Big\{\;\Big[ V_4O_4\left(\bar O_{12}\bar O_{4}+\bar S_{12}\bar S_{4}\right)+O_4V_4\left(\bar V_{12}\bar V_{4}+\bar C_{12}\bar C_{4}\right)\Big]\nonumber \\
&\;\;\;\;+ \!\Big[ V_4O_4\; N^+\bar q\; \bar O_{12}\bar O_4\;\;\; \;\;\;\;\,+O_4V_4\; (N^-\!+4)\bar q \;  \bar O_{12}\bar O_4 \Big]\;\Big\}+\cdots\, ,
\end{align}
where the contributions containing massless states are written explicitly. The left part of the second line involves invariant characters $O_4$ and $\bar O_4,\bar S_4$ under  the $\Z_2$ twist generator, while the right part contains even combinations of odd characters $V_4$ and $\bar V_4,\bar C_4$. Similarly, the left part of the third line involves invariant contributions $O_4$  and $N^+\bar q\, \bar O_4$, while the right part contains even combinations of odd contributions $V_4$ and $(N^-\!+4)\bar q\, \bar O_4$. In total, the massless untwisted degrees of freedom are bosons organized as follows,
\begin{align}
\label{uspec}
\mbox{Bosons in\;\;\;\;\;}&\; [4]_{\psi^{2,3,4,5}_{-1/2}}\otimes \left([4]_{\bar X^{2,3,4,5}_{-1}}\oplus [N^+]\oplus [133]_{E_7}\oplus [3]_{SU(2)}\oplus [248]_{E_8}\right)\nonumber\\
\oplus&\; [4]_{\psi^{6,7,8,9}_{-1/2}}\otimes \left([4]_{\bar X^{6,7,8,9}_{-1}}\oplus [N^-]\oplus [56]_{E_7}\otimes [2]_{SU(2)}\right)\!,
\end{align}
where the right-moving degrees of freedom $[4]_{\bar X^{2,3,4,5}_{-1}}$ arise from $O_{2,2}^{({\rm n.c.})}O_{2,2}\big[{}^0_0\big]$. In our notations, $\psi_{-1/2}^i$ and $\bar X_{-1}^i$, $i\in\{2,\dots, 9\}$ are left-moving fermionic and right-moving bosonic oscillators.  

To show the existence of the $E_7\times SU(2)$ symmetry, we expand the $\bar\theta/\bar\eta$ functions of the~$SO(12)$ affine characters 
\begin{align}
\label{o12}
\bar O_{12}&={1\over \bar q^{6/24}}\Big(1+6\bar q+\!\!\!\sum_{\epsilon_{10},\epsilon_{11}=\pm 1}\!\!\!\bar q^{{2\over 4}[\epsilon_{10}^2+\epsilon_{11}^2+0^4]}+\mbox{\footnotesize  permut. $10,11\to i\neq j\in\{10,\dots, 15\}$}+\O(\bar q^2)\Big),\nonumber \\
\bar V_{12}&={1\over \bar q^{6/24}}\Big(\sum_{\epsilon_{10}=\pm 1}\bar q^{{2\over 4}[\epsilon_{10}^2+0^5]}+\mbox{\footnotesize  permut. $10\to i\in\{10,\dots, 15\}$}+\O(\bar q^{3/2})\Big),\nonumber\\
 \bar S_{12}&={1\over \bar q^{6/24}} \!\!\!\!\!\! \!\!\sum_{\overset{\scriptstyle\epsilon_{10},\dots,\epsilon_{15}=\pm1}{\scriptstyle\epsilon_{10}\cdots\epsilon_{15}=1}}   \!\!\!\!\!\!\bar q^{{2\over 4}\left[({\epsilon_{10}\over 2})^2+\cdots+({\epsilon_{15}\over 2})^2\right]} (1+\O(\bar q)) ,\, \bar C_{12}={1\over \bar q^{6/24}} \!\!\!\!\!\! \!\!\sum_{\overset{\scriptstyle\epsilon_{10},\dots,\epsilon_{15}=\pm1}{\scriptstyle\epsilon_{10}\cdots\epsilon_{15}=-1}}  \!\!\!\!\!\!\bar q^{{2\over 4}\left[({\epsilon_{10}\over 2})^2+\cdots+({\epsilon_{15}\over 2})^2\right]} (1+\O(\bar q)) ,
\end{align}
where $0^k$ denotes a sum of $k$ consecutive $0$'s. 
The brackets in the exponents of $\bar q$ are squares of roots and weights \ie charge 6-vectors under the $U(1)^6$ Cartan generators \cite{Lust-Theisen}. If the~$SO(4)$ affine characters can be expanded in a similar manner in terms of root or weight 2-vectors   
\begin{align}
\bar O_{4}&={1\over \bar q^{2/24}}\Big(1+2\bar q+\sum_{\epsilon,\epsilon'=\pm 1}\bar q^{{2\over 4}[\epsilon^2+\epsilon^{\prime 2}]}+\O(\bar q^2)\Big) ,\nonumber\\
\bar V_{4}&={1\over \bar q^{2/24}}\Big(\sum_{\epsilon=\pm 1}\bar q^{{2\over 4}[\epsilon^2+0]}+\sum_{\epsilon'=\pm 1}\bar q^{{2\over 4}[0+\epsilon^{\prime 2}]}+\O(\bar q^{3/2})\Big),\nonumber \\
\bar S_{4}&={1\over \bar q^{2/24}} \sum_{\overset{\scriptstyle\epsilon,\epsilon'=\pm1}{\scriptstyle\epsilon\epsilon'=1}} \bar q^{{2\over 4}\left[({\epsilon\over 2})^2+({\epsilon'\over 2})^2\right]}\, (1+\O(\bar q))\, ,\quad  \bar C_{4}={1\over \bar q^{2/24}} \sum_{\overset{\scriptstyle\epsilon,\epsilon'=\pm1}{\scriptstyle\epsilon\epsilon'=-1}} \bar q^{{2\over 4}\left[({\epsilon\over 2})^2+({\epsilon'\over 2})^2\right]}\, (1+\O(\bar q))\, ,
\end{align}
it is relevant for our purpose to rotate the orthogonal basis of the 2-dimensional Cartan subalgebra through an angle $\pi/4$, thus interpreting $SO(4)$ as $SU(2)\times SU(2)$ :
\begin{align}
\label{o4}
\bar O_{4}&={1\over \bar q^{2/24}}\Big(1+2\bar q+\sum_{\epsilon_{16}=\pm 1}\bar q^{{2\over 4}[(\epsilon_{16}\sqrt{2})^2+0]}+\sum_{\epsilon_{17}=\pm 1}\bar q^{{2\over 4}[0+(\epsilon_{17}\sqrt{2})^2]}+\O(\bar q^2)\Big) ,\nonumber\\
\bar V_{4}&={1\over \bar q^{2/24}}\Big( \!\sum_{\epsilon_{16},\epsilon_{17}=\pm1} \!\!\!\bar q^{{2\over 4}\left[(\epsilon_{16}{\sqrt{2}\over 2})^2+(\epsilon_{17}{\sqrt{2}\over 2})^2\right]}+\O(\bar q^{3/2})\Big),\nonumber \\
\bar S_{4}&={1\over \bar q^{2/24}} \sum_{\epsilon_{16}=\pm1} \bar q^{{2\over 4}\left[(\epsilon_{16}{\sqrt{2}\over 2})^2+0\right]}\, (1+\O(\bar q))\, ,\quad \bar C_{4}={1\over \bar q^{2/24}} \sum_{\epsilon_{17}=\pm1} \bar q^{{2\over 4}\left[0+(\epsilon_{17}{\sqrt{2}\over 2})^2\right]}\, (1+\O(\bar q))\, .
\end{align}
Using Eqs (\ref{o12}) and (\ref{o4}), one finds the $\bar O_{12}\bar O_{4}+\bar S_{12}\bar S_{4}$ characters in Eq. (\ref{part}) contain a total of $133+3$ right-moving massless degrees of freedom, whose charge 8-vectors are the roots of $E_7\times SU(2)$ \cite{Lust-Theisen}. Similarly, the $\bar V_{12}\bar V_{4}+\bar C_{12}\bar C_{4}$ characters contain  $56\times 2$ massless states, whose weight 8-vectors are those of the $ [56]_{E_7}\otimes [2]_{SU(2)}$ representation.

To summarize, the massless untwisted sector is the bosonic part of $\N=2$ supermultiplets : 1 gravity multiplet (graviton,  graviphoton), 1 tensor multiplet (antisymmetric tensor, dilaton, gauge boson), 1 vector multiplet (gauge boson, complex scalar) in the adjoint representation of $U(1)^2\times E_7\times SU(2)\times E_8$, 1 hypermultiplet (2 complex scalars) in the  $[56]_{E_7}\otimes [2]_{SU(2)}$~\cite{Walton:1987bu}. The remaining part of the massless untwisted spectrum, 
\begin{align}
\label{uspec'}
\mbox{Bosons in\;\;\;\;\;}&\; [4]_{\psi^{2,3,4,5}_{-1/2}}\otimes [N^+]\oplus [4]_{\psi^{6,7,8,9}_{-1/2}}\otimes \left([4]_{\bar X^{6,7,8,9}_{-1}}\oplus [N^-]\right)\!,
\end{align}
amounts to the bosonic degrees of freedom of $N^+$ vector multiplets and $4+N^-$ hypermultiplets. If for $N^\pm=0$ this yields 4 neutral hypermultiplets, we are going to see that non-vanishing $N^\pm$ are required for the model to develop a super no-scale structure, which gives rise to enhanced gauge theories with charged scalars. 

\vspace{.2cm}
{\large \em \noindent The twisted sector}

\noindent In order to find the massless twisted sector, we expand
\be
Z_{4,4}\big[{}^{\, 1}_G\big]\!={16\, q^{1/4}\bar q^{1/4}\over q^{4/24}\bar q^{4/24}}\Big(1+(-1)^G4\bar q^{1/2}+\O(\bar q)\Big)\big(1+\O(q^{1/2})\big)
\ee
and write
\begin{align}
{1\over 2}\left(Z\big[{}^{1}_0\big]\!+ Z\big[{}^{1}_1\big]\right)\!=-O_{2,2}^{({\rm n.c.})}\;&O_{2,2}\big[{}^0_0\big]\; {16\, q^{1/4}\bar q^{1/4}\over q^{4/24}\bar q^{4/24}}\; \bar O_{0,8}^{(E_8)}\; \nonumber \\
\times \; & C_4O_4\, \Big(\bar V_{12}\bar S_{4}+\bar C_{12}\bar O_{4}+4\bar q^{1/2}\; \bar O_{12}\bar C_4\Big)+\cdots\, ,
\end{align}
where the ellipsis contain massive contributions only. In the second line, the characters $O_4$ and $\bar S_4,\bar O_4$ are even under the $\Z_2$ twist, as is also the case for the combination $\bar q^{1/2}\bar C_4$. Using Eqs (\ref{o12}) and (\ref{o4}), the weight 8-vectors of the massless states arising from $\bar V_{12}\bar S_{4}+\bar C_{12}\bar O_{4}$ and  $\bar O_{12}\bar C_4$ have vanishing last entry or vanishing 7 first entries, respectively. These charges of the twisted massless  spectrum yield  
\be
\label{ft}
\mbox{Fermions in \;\;\;\;}32\,[56]_{E_7}\oplus 128\, [2]_{SU(2)}\, ,
\ee
which correspond to  the fermionic parts of 8 hypermultiplets (2 Weyl fermions) in the $[56]_{E_7}$ and 32 hypermultiplets in the $[2]_{SU(2)}$ \cite{Walton:1987bu}. The number of twisted massless fermionic degrees of freedom  is thus 
\be
\label{nf}
\nF^{\rm t}=4\times 512\, .
\ee

\vspace{.2cm}
{\large \em \noindent The super no-scale condition}

\noindent For the $\N=2\to 0$ model to develop a super no-scale structure, we require $\nB^{\rm u}$ given in Eq.~(\ref{nb}) to equal  $\nF^{\rm t}$, \ie an enhanced gauge symmetry in the parent $\N=4\to 0$ model such that $\dim G^{(T^4)}=12$. The rank of $ G^{(T^4)}$ being 4, we must have 
\be
\label{sol}
(a) \;\;G^{(T^4)}=SU(2)_{\rm en}^4\qquad \mbox{or} \qquad (b) \;\;G^{(T^4)}=SU(3)_{\rm en}\times SU(2)_{\rm en}\times U(1)\, .
\ee
These solutions are realized when $T^4$ is factorized as $T^2\times T^2$. Thus, we can define $T_2,U_2$ and $T_3,U_3$ to be the \Ka and complex structure moduli of these second and third $T^2$'s, in a way similar to those defined in Eq. (\ref{TU}) for the first one. Modulo T-duality, the solution~$(a)$ is obtained at the self-dual point  $T_2=U_2=T_3=U_3=i$, while the solution $(b)$ occurs at $T_2=U_2=e^{i\pi/3}$, $T_3=U_3$ arbitrary. 

\vspace{.2cm}
{\large \em \noindent The $4N^+$ and $4N^-$  untwisted bosonic states}

\noindent Since the above solutions yield $N^+=N^-=4$, we have to  describe the representations of the spectrum given in Eq. (\ref{uspec'}).

$\bullet$  In the solution $(a)$, for each direction $i\in\{6,7,8,9\}$, the right-moving momentum $p^i_R={1\over \sqrt{2}}(m_i/R_i-n_iR_i)$ where $R_i=1$ is a charge under an enhanced $SU(2)_{\rm en}$ gauge symmetry in the parent $\N=4\to 0$ model. At the massless level, the states 
$m_i=-n_i=\pm 1$ (at right-moving oscillator level 0) and $m_i=-n_i=0$ (at right-moving oscillator level 1) have charges~$\pm \sqrt{2}$, 0 \ie realize an adjoint representation, $[3]_{SU(2)_{\rm en}}$. Therefore, in the descendant orbifold model, the contributions $N^+\bar q$ and $(N^-\!+4)\bar q$ in Eq. (\ref{part}), which are even and odd under the $\Z_2$ generator, are in 4 copies of the same representation (one for each $SU(2)_{\rm en}$ factor). In fact, $V_4O_4 \, N^+\bar q$ is the contribution of the bosonic parts of  4 vector multiplets,  thus realizing an enhanced $U(1)_{\rm en}^4$ gauge symmetry. Moreover, $O_4V_4\, (N^-\!+4)\bar q$ corresponds to the bosonic parts of 4 pairs of hypermultiplets charged under one of the Abelian gauge factors, with charges  $\pm\sqrt{2}$.

In total, the massless spectrum arising in the solution $(a)$ from the enhanced symmetry of the $T^4$ lattice amounts to the bosonic parts of 4 copies of the following set of $\N=2$ supermultiplets~:  1 Abelian vector multiplet, 1 hypermultiplet of charge $+\sqrt{2}$, 1 hypermultiplet of charge $-\sqrt{2}$. The full gauge symmetry of the $\N=2\to 0$ model is 
\be
\label{gga}
U(1)^2_{\rm grav,ten}\times U(1)^2\times U(1)^4_{\rm en}\times E_7\times SU(2)\times E_8\, ,
\ee
where $U(1)^2_{\rm grav,ten}$ refers to the gravity and tensor multiplets gauge bosons. 

$\bullet$  In the solution $(b)$, the $T^4$ lattice induces an enhanced $SU(3)_{\rm en}\times SU(2)_{\rm en}\times U(1)$ gauge symmetry in the parent $\N=4\to 0$ model. Organizing the following contributions of Eq.~(\ref{part}) as 
\begin{align}
&V_4O_4\; N^+\bar q=V_4O_4\,(3\bar q+1\bar q+0\bar q)\, , \nonumber  \\
&O_4V_4\; (N^-\!+4)\bar q= O_4V_4\, \big((3+2)\bar q+(1+1)\bar q+(0+1)\bar q\big)\, ,
\label{sp}
\end{align}
we can write $3\bar q$ and $(3+2)\bar q$ in terms of $SU(3)_{\rm en}$ roots and  Cartan  generators,
\begin{align}
&\bar q^{{2\over 4}[(\epsilon_6 \sqrt{2})^2+0]}\; , \; \bar q^{1+0}\quad\;\;  \qquad \Longrightarrow \quad [3]_{SU(2)_{\rm en}}\nonumber \\
&\bar q^{{2\over 4}\left[(\epsilon_6{\sqrt{2}\over 2})^2+(\epsilon_7{\sqrt{2}\over 2})^2\right]}\; , \; \bar q^{0+1}\quad \Longrightarrow \quad 2 [2]_{SU(2)_{\rm en}}\oplus[1]_{SU(2)_{\rm en}}\, ,
\end{align}
where $\epsilon_6,\epsilon_7=\pm 1$. This shows the existence of an $SU(2)_{\rm en}$ gauge symmetry in the descendent $\N=2\to 0$ model, coupled to 2 copies (for $\epsilon_7=\pm1$) of scalar fields in the fundamental representation, and neutral scalars. Moreover, as was the case in the solution $(a)$, the terms $1\bar q$ and $(1+1)\bar q$  in Eq (\ref{sp}) lead an $U(1)_{\rm en}$ gauge symmetry coupled to fields of charges~$\pm \sqrt{2}$. Finally, the contribution $(0+1) \bar q$ corresponds to neutral scalar fields. 

In total, the massless spectrum arising in the solution $(b)$ from the enhanced symmetry of the $T^4$ lattice amounts to the bosonic parts of $\N=2$ supermultiplets~:  1 vector multiplet in the adjoint representation of $SU(2)_{\rm en}\times U(1)_{\rm en}$, 2 hypermultiplets in the $[2]_{SU(2)_{\rm en}}$, 1~hypermultiplet of charge $\sqrt{2}$ under $U(1)_{\rm en}$, 1 hypermultiplet of charge $-\sqrt{2}$ under $U(1)_{\rm en}$ and 2 neutral hypermultiplets. The full gauge symmetry is therefore 
\be
\label{gg}
U(1)^2_{\rm grav,ten}\times U(1)^2\times SU(2)_{\rm en}\times U(1)_{\rm en}\times E_7\times SU(2)\times E_8\, .
\ee

\vspace{.2cm}
{\large \em \noindent The would-be tachyons}

\noindent It is well known that the insertion of the spin structure dependent sign $S$, Eq. (\ref{S}), in a $\mbox{1-loop}$ partition function can yield a Hegedorn-like instability when $\m=\O(\Ms)$. Actually, coupling the spin structure $a,b$ with a lattice shift $h,g$ along an Euclidean time circle instead of an internal spatial direction amounts to switching on finite temperature $T$ rather than supersymmetry breaking, which leads to a Hagedorn divergence for $T=\O(\Ms)$ \cite{Kounnas-Rostand,Hage}. 

For instance, considering the partition function as given in Eq. (\ref{Z}) but with sign insertion $S$ only, untwisted scalars ($H=0$) with momentum and winding numbers $2k_4+g=-(2n_4+h)=\pm 1$ become tachyonic when $\m$ approaches $\Ms$. However, with $S$ and $S'$ inserted, we have shown that the model does not develop physical, \ie level matched, tachyons.  This can be seen in  Eq.~(\ref{00}), reminding that  $Z\big[{}^{0}_1\big]\!=0$ and that tachyons may only occur in the untwisted sector, $H=0$. This different behaviour  may be surprising, since it seems that we have show that the light spectrum in the untwisted sector is not modified by the insertion of $S'$ (!)

The resolution of this puzzle comes from the fact that the two models share their light untwisted states in the sector $h=0$ only. Thus, they have identical  light spectra if $\m$ is low enough for the sector $h=1$ not to be light. Since the tachyons may only occur when $\m =\O(\Ms)$ and have $h=1$,  their presence may be affected by the insertion of $S'$.
To see this is  the case, we observe  that in a sector  $h=1$, $H=0$, we have $S'=(-1)^G$. Denoting~$\boldsymbol{G}_S$ the $\Z_2$ generator in a model where the supersymmetry breaking is implemented with $S$ only, the projector on $\Z_2$-invariant states becomes, once we insert $S'$,
\be
{1\over 2}\sum_{G=0}^1\boldsymbol{G}_S^G={1+\boldsymbol{G}_S\over 2}\quad \longrightarrow \quad {1\over 2}\sum_{G=0}^1(-1)^G\boldsymbol{G}_S^G={1-\boldsymbol{G}_S\over 2}\, .
\ee
Thus, the tachyons surviving the $\Z_2$-projection with $S$ alone are projected out when $S'$ is included.\footnote{In the last line of Eq. (\ref{Z0}), the sign $(-1)^G$ dresses the characters $O_4,S_4$ and not $V_4,C_4$, as is the case in the model where only $S$ is inserted.}  
This statement is true even when marginal deformations  are switched on, provided they are small enough. However, when $\m$ is of order $\Ms$ and $\O(1)$ moduli are turned on, most models can develop tachyonic instabilities at tree level (see Ref. \cite{Angelantonj:2006ut} for a counterexample).


\section{Moduli stability}
\label{defor}

At low enough $\m$, when a no-scale model sits at a point in moduli space where it develops a super no-scale structure, the term proportional to $\m^4$ in the 1-loop effective potential drops. At this order in perturbation theory, up to exponentially small corrections, the no-scale modulus $\m$ has no tadpole and remains a flat direction. However, other marginal deformations of the classical theory exist. A important issue is that the gauge theory described by the undeformed super no-scale background may admit various branches : A~Coulomb phase, a Higgs phase, and in general numerous mixed Coulomb/Higgs phases. In this case, the marginal deformation can be along any branch. 

The goal of the present section is to turn on all moduli that parameterize some of these phases. In the class of models defined in Sect.~\ref{2->0}, we consider in details the deformations along the pure Coulomb branch, as well as along some of the mixed branches. In each phase, we compute the corresponding expression of the 1-loop effective potential at quadratic order in moduli fields, and conclude on the possible destabilization of the classical background in these directions.
In the following, after stating some generalities about the derivation of the effective potential, we analyse the case of the background~$(a)$. 

Notice that before deformation of a super no-scale background, there are $n_B=n_F$ bosonic and fermionic massless degrees of freedom and no mass scale below $\m$. Thus, turning on small but generic deformations of a super no-scale background amounts physically to introducing masses lower than $\m$ to some of these initially massless states.   At such a point in moduli space,  a  general expression for $\Vone^{\N=2\to 0}$ exists for any heterotic no-scale model (super or not) on $T^2\times K3$, when the stringy Scherk-Schwarz mechanism operates along the  direction $X^4$ of the large $T^2$. It takes the form
\be
\label{Vi}
\V_{\mbox{\scriptsize 1-loop}}^{\N=2\to 0}=-{M_{\rm s}^4\over (2\pi)^4}\sum_{s_0=1}^{n_B+n_F}(-1)^{F_0}\int_0^{+\infty}{d\tau_2\over 2\tau_2^3}\sum_{m_4,m_5}(-1)^{m_4}\,e^{-\pi\tau_2 M_L^{2}/M_{\rm s}^2}+\O\!\left({c^2\Ms^4\over \Im T_1}\, e^{-c\sqrt{\Im T_1}}\right)\!,
\ee
where $s_0$ denotes one of the $n_B+n_F$  bosonic or fermionic degrees of freedom of mass below $\m$ and fermion number $F_0$. The dominant contribution comes from the Kaluza-Klein towers of modes arising from the large $T^2$ and based on the states $s_0$. $M_L$ is the left-moving mass of each Kaluza-Klein state and depends on the moduli deformations. The exponentially suppressed corrections involve a moduli-dependent positive quantity $c$ defined as  $M_{\rm high}/\m=c\Ms/\m\simeq c\sqrt{\Im T_1}$, where $M_{\rm high}$ is the lowest mass scale above $\m$ in the spectrum, which in practice can be very high. The justification of Eq. (\ref{Vi}) can be found in Ref. \cite{SNSM} but can be summarized as follows : 

$\bullet$ The existence of an infinite tower of Kaluza-Klein states associated to the direction~$X^4$ involved in the supersymmetry breaking implies the partition function to be integrable over the full upper half strip, $-1/2<\tau_1<1/2$, $\tau_2>0$. No ultraviolet divergence occurs as~$\tau_2\to 0$. 

$\bullet$ Moreover, when the volume of this compact direction is large, compared to the string scale,  the integral over the region $\tau_2<\sqrt{3}/2$ of the strip  is exponentially suppressed. Therefore, when $\m$ is lower than $\Ms$, we can replace  up to exponentially suppressed terms the domain of integration over  the fundamental domain of $SL(2,\Z)$ by the upper half strip.   

$\bullet$ The non-level matched states are projected out of this integral and only  the physical ones remain.

$\bullet$ Among them, the degrees of freedom heavier than the lowest mass scale $M_{\rm high}$ above $\m$ yield exponentially suppressed contributions, compared to those arising from the $T^2$ Kaluza-Klein modes based on the states $s_0$ whose masses are below $\m$. In particular, the oscillator modes at mass level $\Ms$ and the winding states along $T^2$ are suppressed.  

$\bullet$ Finally, each boson (fermion)  $s_0$ is accompanied by an infinite tower of Kaluza-Klein bosons (fermions) with momenta $m_4=2k_4$ and $m_5$, as well as another tower of Kaluza-Klein fermions (bosons)  with $m_4=2k_4+1$ and $m_5$. Thus, the fermion number of any  Kaluza-Klein state of momentum $m_4$ is the parity of $F_0+m_4$, which justifies the sign $(-1)^{F_0+m_4}$ in Eq.~(\ref{Vi}).

From now on, we consider in great details the case of  the background ($a$), whose gauge  symmetry group is given in Eq. (\ref{gga}). This model illustrates the class of theories presented in Sect. \ref{2->0}, where all twisted moduli present at the exact $\N=2$ level are lifted classically. The scalar fields that may parameterize marginal deformations along any phase of the gauge theory can be listed from the massless untwisted spectrum of Eq. (\ref{uspec}), namely :  

$\bullet$ The internal metric and antisymmetric tensor of the large $T^2$. They can be expressed in terms of those of the initial background we denote with upper indices~``$(a)$"  and $2\times 2$ deformations,
\be
\label{bg}
(B+G)_{\alpha\beta}=(B^{(a)}+G^{(a)})_{\alpha\beta}+ Y_{\alpha\beta}\, , \qquad \alpha,\beta\in\{4,5\}\, .
\ee
They correspond to the degrees of freedom $[4]_{\psi^{4,5}_{-1/2}}\otimes [4]_{\bar X^{4,5}_{-1}}$ \ie $T_1,U_1$, which parameterize the Coulomb branch of the $U(1)^2$ gauge symmetry generated by the $T^2$ lattice. 
 
$\bullet$ The internal metric and antisymmetric tensor of $T^4$. They can be expressed in terms of the metric of the Cartesian product of four circles of unit radii and $4\times 4$ deformations, 
\be
\label{bg2}
(B+G)_{ij}\;=\delta_{ij}+\sqrt{2}\, Y_{ij}\, , \qquad  i,j\in\{6,7,8 ,9\}\, ,
\ee
corresponding to the degrees of freedom $[4]_{\psi^{6,7,8,9}_{-1/2}}\otimes [4]_{\bar X^{6,7,8,9}_{-1}}$. However, we showed above Eq.~(\ref{gga}) that for each $i\in\{6,7,8,9\}$, the quaternion $Y_{ji}$, $j\in\{6,7,8,9\}$, is of charge~$-\sqrt{2}$ under the $U(1)_{\rm en}$ factor arising from the direction $X^i$. Therefore, its condensation breaks spontaneously this $U(1)_{\rm en}$ and we conclude that, together, the $Y_{ji}$, $j,i\in\{6,7,8,9\}$, parameterize the pure Higgs branch of the $U(1)^4_{\rm en}$ gauge theory.  

$\bullet$ The $E_7\times SU(2)\times E_8$ Wilson lines along $T^2$. From the point of view of the parent $\N=4\to 0$ model, they are the $2\times 16$ Wilson lines of $E_8\times E_8$ that survive the $\Z_2$ projection,  
\be
\label{wl}
Y_{\alpha \I}\, , \qquad \alpha\in\{4,5\}\, , \; \I\in\{10,\dots,25\}\, . 
\ee
They are denoted  $[4]_{\psi^{4,5}_{-1/2}}\otimes \left.\left([133]_{E_7}\oplus [3]_{SU(2)}\oplus [248]_{E_8}\right)\!\right\abs_{\rm Cartan}$ in Eq.~(\ref{uspec}) and parameterize the pure Coulomb branch of the  $E_7\times SU(2)\times E_8$ gauge theory.

$\bullet$ The Wilson lines along $T^2$ of the gauge group factor of dimension $N^+$.  These states, denoted  $[4]_{\psi^{4,5}_{-1/2}}\otimes [N^+]\abs_{\rm Cartan}$, are rank[$U(1)^4_{\rm en}]=4$ complex scalars, which parameterize the pure Coulomb phase of the $U(1)^4_{\rm en}$ gauge theory.

$\bullet$ The 4 quaternionic scalars $[4]_{\psi^{6,7,8,9}_{-1/2}}\otimes [N^-]$. Each of them has charge $\sqrt{2}$ under one of the $U(1)_{\rm en}$ factors.  They are on equal footing with the quaternions of opposite charges $Y_{ji}$, $j,i\in\{6,7,8,9\}$, and it is matter of convention to use the former or the latter to parameterize the pure Higgs branch of the  $U(1)_{\rm en}^4$ gauge theory. 

$\bullet$ The quaternionic scalars $[4]_{\psi^{6,7,8,9}_{-1/2}}\otimes \left([56]_{E_7}\otimes [2]_{SU(2)}\right)$. They are in the bifundamental representation and their condensation can break spontaneously $E_7\times SU(2)$ to various subgroups of  ranks $r<8$. In each case, the possibility to explore a Coulomb phase of complex dimension $r$ remains, which yields mixed Coulomb/Higgs branches of the $E_7\times SU(2)$ gauge theory. 

In the following, we will not evaluate the 1-loop effective potential in the mixed phases reached \via condensation of degrees of freedom of the quaternionic $[56]_{E_7}\otimes [2]_{SU(2)}$. However, we will explore  in details all branches of the $U(1)^2_{\rm grav,ten}\times U(1)^2\times U(1)^4_{\rm en}\times E_7\times SU(2)\times E_8$ gauge theory, where the $E_7\times SU(2)$ gauge symmetry is restricted to its pure Coulomb phase.

\vspace{.2cm}
{\large \em \noindent Contribution of the untwisted states}

\noindent We first consider the branch of the gauge theory where $U(1)_{\rm en}^4$ is in its pure Higgs phase and $U(1)^2_{\rm grav,ten}\times U(1)^2\times E_7\times SU(2)\times E_8$ in its pure Coulomb phase. In this case, the allowed background deformations are those given in Eqs  (\ref{bg})--(\ref{wl}). To compute the contribution $\Vone^{\N=2\to 0, \rm u}$  of the 1-loop effective potential arising from the Kaluza-Klein towers of untwisted states, we can follow two strategies. 
We can consider all untwisted massless states $s_0$ present in the $\N=2\to 0$ background $(a)$, compute the deformed masses $M_L$ of their Kaluza-Klein modes along $T^2$, and apply Eq. (\ref{Vi}). If this is what we will do later in this section, we find convenient to start with the second approach. 

Thanks to the identity~(\ref{id}), which is valid even in the deformed background, the sector $h=0$ in $Z\big[{}^{0}_1\big]$ vanishes. Thus, we can write 
\be
\Vone^{\N=2\to 0,\rm u}={1\over 2}\, \Vone^{\N=4\to 0} +\O\!\left(\Ms^4\, e^{-\Im T_1}\right)\!,
\ee
where $\Vone^{\N=4\to 0}$ is the potential of the parent $\N=4\to 0$ model, while  the suppressed terms arise from winding modes along $X^4$. In order to compute the effective potential $\Vone^{\N=4\to 0}$, the massless states $s_0$ to be considered are those of the parent $\N=4\to 0$ model, which  are charged under $SU(2)_{\rm en}^4\times E_8\times E_8$. The left-moving squared  masses of their Kaluza-Klein modes along $X^{4,5}$ can be written as    \cite{421}, 
\be
\label{ML}
M_L^{2}= M_{\rm s}^2 \, P_I\, G^{-1}_{IJ}P_J\, , 
\ee
in terms of generalized momenta
\be
\label{P}
P_I=m_I+Y_{I\I}\, Q_\I+{1\over 2}Y_{I\I}\, Y_{J\I}\, n_J+(G+B)_{IJ}\, n_J\, ,
\ee
where implicit sums over $I,J\in\{4,\dots,9\}$ and $\I\in\{10,\dots,25\}$ are understood. In the above expression, $m_I,n_I$ are the momenta and winding numbers along the direction $X^I$ (we have $n_4=n_5=0$), while $Q_\I$ are the roots of $E_8\times E_8$. Actually, this formula is valid for arbitrary $T^6$ metric and antisymmetric tensor, $(G+B)_{IJ}$, as well as arbitrary $E_8\times E_8$ Wilson lines along~$T^6$, $Y_{I\I}$. Therefore, the expression of  $\V^{\N=2\to 0,\rm u}_{\mbox{\scriptsize 1-loop}}$ we are interested in is that of $\V^{\N=4\to 0}_{\mbox{\scriptsize 1-loop}}$, with a restricted set of deformations. 

The states $s_0$ of the parent $\N=4\to 0$ model are  : 

$\bullet$ The 8 bosons of the $\N=4$ vector multiplets charged under the $SU(2)_{\rm en}$ gauge factor arising from the direction $X^i$, for $i\in \{6,7,8,9\}$. As said before, their right-moving momentum $p^i_R={1\over \sqrt{2}}(m_i/R_i-n_iR_i)$, with $R_i=1$ and $m_i=-n_i=\pm 1$, are the roots $Q_i=\pm \sqrt{2}$ of $SU(2)_{\rm en}$. Their other quantum numbers are trivial,  namely $m_j=n_j=Q_\I=0$,  $j\in\{6,7,8,9\}$, $j\neq i$, $\I\in\{10,\dots, 25\}$.

$\bullet$ The 8 bosons of the $\N=4$ vector multiplets charged under the first (\ie $\kappa=0$) or second (\ie $\kappa=1$) $E_8$ gauge factors. The charges $Q_{\I+8\kappa}$, $\I\in \{10,\dots, 17\}$, are roots of the corresponding $E_8$, 
\be
Q_{\I+8\kappa}=\left\{
\begin{array}{ll}(\pm1,\pm1,0,0,0,0) & \mbox{or permutations,}\\\mbox{\!\!\!or}\\
(\pm1,\pm1,\pm1,\pm1,\pm1,\pm1,\pm1,\pm1)  & \mbox{with even number of $-1$'s.}
\end{array}
\right.
\ee
Their other quantum numbers vanish,  $m_i=n_i=Q_{\I+8(1-\kappa)}=0$,  $i\in\{6,7,8,9\}$. 

$\bullet$ The $8\times 22$ bosons of the $\N=4$ vector multiplets in the Cartan subalgebra of  $U(1)^2\times SU(2)_{\rm en}^4\times E_8\times E_8$, together with the $8\times 2$ ones of the $\N=4$ supergravity multiplet. They  have vanishing charges,  $m_i=n_i=Q_\I=0$,  $i\in\{6,7,8,9\}$, $\I\in \{10,\dots,25\}$.

Following the steps of Ref. \cite{SNSM}, it is then possible to compute for each state $s_0$ the squared masses $M_L^2$ of the $T^2$ Kaluza-Klein states and evaluate $\V^{\N=2\to 0,\rm u}_{\mbox{\scriptsize 1-loop}}$ at second order in $Y_{ji}$, $Y_{\alpha\I}$, $j,i\in\{6,7,8,9\}$, $\alpha\in\{4,5\}$, $\I\in\{10,\cdots,25\}$. The result can be written as
\begin{align}
\label{Vu}
\V^{\N=2\to 0,\rm u}_{\mbox{\scriptsize 1-loop}}= {1\over 2}&\, \Bigg\{ {0-\nB^{\N=4\to 0}\over 16\pi^7}\,{ M_{\rm s}^4\over (\Im T_1)^2}\, E_{(1,0)}(U_1\abs 3,0)+{1\over 16\pi^5}\,{ M_{\rm s}^4\over \Im T_1}\, E_{(1,0)}(U_1\abs 2,0)\; \times 
\desp\nonumber  \\
&\qquad \Bigg[\sum_{i=6}^9 8\; {1\over 2}\!\!\sum_{\underset{\mbox{\scriptsize of $[3]_{SU(2)}$}}{\mbox{\scriptsize roots $Q_i$}}}\; \sum_{j=6}^9\big(Y_{ji}\,Q_i\big)^2\nonumber\tesp\\
& \qquad +8\; {1\over 2}\!\!\!\!\!\sum_{\underset{\mbox{\scriptsize $[248]_{E_8}\oplus[248]_{E_8}$}}{\mbox{\scriptsize roots $Q$ of}}}\!\!\!\!\!\!\left(2\Big\abs \sum_{\I=10}^{25}Y_\I\, Q_\I\Big\abs^2-\rho\Big(\sum_{\I=10}^{25}Y_\I\, Q_\I\Big)^2-\bar\rho\Big(\sum_{\I=10}^{25}\bar Y_\I\, Q_\I \Big)^2\right)\Bigg]\nonumber\\
&\hspace{6.9cm}+\cdots+\O\!\left({c^2\Ms^4\over \Im T_1}\, e^{-c\sqrt{\Im T_1}}\right)\!\Bigg\}\, ,
\end{align}
where $\nB^{\N=4\to 0}=8\times 512$ is the number of massless bosons in the parent $\N=4\to 0$ model before deformation, and the dots stand for higher order corrections in $Y$'s. The $T^2$ moduli $T_1,U_1$ are those defined with the new background, Eq.~(\ref{bg}), while the $E_8\times E_8$ Wilson lines along $T^2$ are redefined in complex notation,
\be
Y_\I={U_1 Y_{4\I}-Y_{5\I}\over \sqrt{\Im T_1\Im U_1}}\, ,\qquad \I\in\{10,\dots,25\}\, . 
\ee 
The Kaluza-Klein towers of states yield shifted complex Eisenstein series of asymmetric integer modulo 2 weights $g_1,g_2$,
\be
E_{(g_1,g_2)}(U\abs  s,k)={\sum_{\tilde m_1,\tilde m_2}}^{\!\!\!\prime}{(\Im U)^s\over \left(\tilde m_1+{g_1\over 2}+(\tilde m_2+{g_2\over 2})U\right)^{s+k}(\tilde m_1+{g_1\over 2}+\left(\tilde m_2+{g_2\over 2})\bar U\right)^{s-k}}
\ee
and we have defined the coefficient
\be
\rho={E_{(1,0)}(U_1\abs 2,1)\over E_{(1,0)}(U_1\abs 2,0)}\, .
\ee
Note that from the point of view of the parent $\N=4\to 0$ theory, all deformations in Eq.~(\ref{Vu}) are Wilson lines. In particular, the $T^4$ metric and antisymmetric tensor deformation $Y_{ji}$, $j,i\in\{6,7,8,9\}$, is the Wilson line along $X^j$ of the $SU(2)_{\rm en}$ factor arising at the self-dual radius $R_i=1$ of the compact direction $X^i$. 
  
\vspace{.2cm}
{\large \em \noindent Contribution of the twisted states}

\noindent To evaluate the contribution of the twisted sector to the 1-loop effective potential, $\V^{\N=2\to 0,\rm t}_{\mbox{\scriptsize 1-loop}}$, the massless states $s_0$ of the undeformed $\N=2\to 0$ background to be considered are given in Eq. (\ref{ft}). Being twisted along $T^4$, they have $m_i=n_i=0$, $i\in\{6,7,8,9\}$. Thus, once the marginal deformations (\ref{bg})--(\ref{wl})  are switched on, their Kaluza-Klein modes along $T^2$ have $P_i=0$ (see Eq. (\ref{P})) and their left-moving squared masses are 
\be
M_L^2=\Ms^2(m_\alpha+\xi_\alpha)G^{-1}_{\alpha\beta}(m_\beta+\xi_\beta) \qquad \where \qquad \xi_\alpha=\sum_{\I=10}^{25}Y_{\alpha\I}\, Q_\I\, ,\quad  \alpha\in\{4,5\}\, .
\ee
In this expression,  the charges $Q_\I$, $\I\in \{10,\dots, 17\}$, are either the weights of the $[56]_{E_7}$,
\begin{align}
Q_\I=\left\{
\begin{array}{ll}\!\big(\!\pm1,0,0,0,0,0,\pm {\sqrt{2}\over 2},0\big) & \mbox{or permutations of the 6 first entries,}\\\mbox{\!\!\!or}\\
(\pm1,\pm1,\pm1,\pm1,\pm1,\pm1,0,0)  & \mbox{with even number of $-1$'s,}
\end{array} 
\right.
\end{align}
or those of the $[2]_{SU(2)}$,
\be
Q_\I=\bigg(0,0,0,0,0,0,0,\pm {\sqrt{2}\over 2}\bigg)\, ,
\ee
while the unbroken  $E_8$ charges are trivial, $Q_\I=0$, $\I\in\{18,\dots, 25\}$. 
Proceeding  as in Ref. \cite{SNSM}, Eq. (\ref{Vi})  yields
\begin{align}
\label{Vt}
\V^{\N=2\to 0,\rm t}_{\mbox{\scriptsize 1-loop}}= &\, {\nF^{\rm t}\over 16\pi^7}\,{ M_{\rm s}^4\over (\Im T_1)^2}\, E_{(1,0)}(U_1\abs 3,0)-{1\over 16\pi^5}\,{ M_{\rm s}^4\over \Im T_1}\, E_{(1,0)}(U_1\abs 2,0)\, \times \desp\nonumber  \\
& \qquad \Bigg[4\times 8\; {1\over 2}\!\sum_{\underset{\mbox{\scriptsize of $[56]_{E_7}$}}{\mbox{\scriptsize weights $Q$}}}\!\left(2\Big\abs \sum_{\I=10}^{16}Y_\I\, Q_\I\Big\abs^2-\rho\Big(\sum_{\I=10}^{16}Y_\I\, Q_\I\Big)^2-\bar\rho\Big(\sum_{\I=10}^{16}\bar Y_\I\, Q_\I \Big)^2\right)\nonumber\tesp\\
& \qquad +4\times 32\; {1\over 2}\!\sum_{\underset{\mbox{\scriptsize of $[2]_{SU(2)}$}}{\mbox{\scriptsize weights $Q$}}}\!\left(2\big\abs Y_{17}\, Q_{17}\big\abs^2-\rho\big(Y_{17}\, Q_{17}\big)^2-\bar\rho\big(\bar Y_{17}\, Q_{17} \big)^2\right)\Bigg]\nonumber\\
&\hspace{6.9cm}+\cdots+\O\!\left({c^2\Ms^4\over \Im T_1}\, e^{-c\sqrt{\Im T_1}}\right) .
\end{align}

\vspace{.2cm}
{\large \em \noindent The total 1-loop effective potential}

\noindent Combining the untwisted and twisted states contributions, Eqs (\ref{Vu}) and (\ref{Vt}), the would-be dominant term proportional to $\Ms^4/(\Im T_1)^2 \propto \m^4$ cancels out, due to the super no-scale condition. In order to simplify the charge-dependent corrections, we use the fact that 
\be
 {1\over 2}\!\!\sum_{\underset{\mbox{\scriptsize of $\cal R$}}{\mbox{\scriptsize weights $Q$}}}\sum_{\I}A_\I Q_\I\, \sum_{\J}B_\J Q_\J=C_{\cal R}\sum_{\I}A_\I B_\I\, , 
\ee
where $\cal R$ is a representation of $SU(2)$, $E_8$ or $E_7$ and the sums over $\I,\J$ run over the corresponding rank,
\be
C_{[3]_{SU(2)}}=2\, , \quad C_{[2]_{SU(2)}}={1\over 2}\, , \quad C_{[248]_{E_8}}=30\, , \quad C_{[56]_{E_7}}=6\, ,\quad C_{[133]_{E_7}}=18\, .
\ee
In total, we obtain in the branch where $U(1)_{\rm en}^4$ is in its pure Higgs phase and $U(1)^2_{\rm grav,ten}\times U(1)^2\times E_7\times SU(2)\times E_8$ in its pure Coulomb phase
\begin{align}
\label{Vtot}
\V_{\mbox{\scriptsize 1-loop}}^{\N=2\to 0}= &\; {1\over 16\pi^5}\,{ M_{\rm s}^4\over \Im T_1}\, E_{(1,0)}(U_1\abs 2,0)\Bigg[ \sum_{i=6}^9 c_{U(1)_{\rm en}} \sum_{j=6}^9 (Y_{ji})^2\desp\nonumber  \\
&+c_{E_7} \sum_{\I=10}^{16}\Big(2\abs Y_\I\abs^2-\rho(Y_\I)^2-\bar\rho(\bar Y_\I)^2\Big)+ c_{SU(2)} \Big(2\abs Y_{17}\abs^2-\rho(Y_{17})^2-\bar\rho(\bar Y_{17})^2\Big)\nonumber\tesp\\
&+c_{E_8} \sum_{\I=18}^{25}\Big(2\abs Y_\I\abs^2-\rho(Y_\I)^2-\bar\rho(\bar Y_\I)^2\Big)\Bigg]\!+\cdots+\O\!\left({c^2\Ms^4\over \Im T_1}\, e^{-c\sqrt{\Im T_1}}\right) ,
\end{align}
where 
\begin{align}
c_{U(1)_{\rm en}}&={8\over 2}\, C_{[3]_{SU(2)}}=8\, , &&  \!\!\!\!\!\!c_{E_7}={8\over 2}\, C_{[248]_{E_8}}-4\times 8\,C_{[56]_{E_7}}=-72\, , \nonumber \\
c_{SU(2)}&={8\over 2}\, C_{[248]_{E_8}}-4\times 32\, C_{[2]_{SU(2)}}=56\, , &&\!\!\!\!\!\!c_{E_8}={8\over 2}\, C_{[248]_{E_8}}=120\, .
\end{align}

As a cross check, we can recompute the dependance on the $E_7\times SU(2)\times E_8$ Wilson lines of the untwisted sector contribution $\Vone^{\N=2\to 0,\rm u}$ to the potential.  As announced before, this can be done directly from the point of view of the untwisted spectrum of the $\N=2\to 0$ model. From Eq. (\ref{uspec}), we see that the states $s_0$ charged under $E_7$ are the bosons of an $\N=2$ vector multiplet in the $[133]_{E_7}$ and those of 2 hypermultiplets in the $[56]_{E_7}$. Similarly, charged under $SU(2)$, we have the bosons of an $\N=2$ vector multiplet in the $[3]_{SU(2)}$ and those of 56 hypermultiplets in the $[2]_{SU(2)}$. Finally, charged under $E_8$, we  have the bosons of an $\N=2$ vector multiplet in the $[248]_{E_8}$. Due to the identities
\be
4(C_{[133]_{E_7}}+2 C_{[56]_{E_7}})= {8\over 2}\, C_{[248]_{E_8}} , \, 4(C_{[3]_{SU(2)}}+56 C_{[2]_{SU(2)}})= {8\over 2}\,C_{[248]_{E_8}}, \, 4C_{[248]_{E_8}}= {8\over 2}\,C_{[248]_{E_8}} ,
\ee
we find perfect agreement with the analysis based on the parent $E_8\times E_8$ theory. 

We said before that from the $\N=4\to 0$ viewpoint, the scalars $Y_{ji}$, $j\in\{6,7,8,9\}$, are four out of six degrees of freedom parameterizing the Coulomb branch of the $SU(2)_{\rm en}$ gauge factor arising from the direction $X^i$,  $i\in\{6,7,8,9\}$. In the descendent $\N=2\to 0$ model, we have also shown that $Y_{ji}$, $j\in\{6,7,8,9\}$, are instead a quaternion of charge $-\sqrt{2}$ parameterizing the Higgs branch of the  $U(1)_{\rm en}$ gauge factor arising from the direction $X^i$. To see the two viewpoints are consistent, we note that if a $Y_{ji}$ condenses in the parent theory, an $SU(2)_{\rm en}$ is spontaneously broken to $U(1)$ and  $N^+=N^-$ decrease by one unit in the descendent orbifold model. Physically, this means that the bosonic part of an $U(1)_{\rm en}$ $\N=2$ vector multiplet has combined with the quaternion of charge $\sqrt{2}$ to become the bosonic part of a long massive $\N=2$ vector multiplet.\footnote{An $SU(2)_R$ triplet of $U(1)_{\rm en}$ D-terms conditions fix 3 components of the quaternion of charge $\sqrt{2}$. The  remaining component of the quaternion is gauged away by the residual global  $U(1)_{\rm en}$ symmetry. It is the would-be Goldstone boson ``eaten" by the massive gauge boson.} The $U(1)_{\rm en}$ gauge theory is in its Higgs phase, where the only massless states are the neutral quaternion $Y_{ji}$, $j\in\{6,7,8,9\}$. 

Next, we consider the branch of the gauge theory where $U(1)_{\rm en}^4$ as well as $U(1)^2_{\rm grav,ten}\times U(1)^2\times E_7\times SU(2)\times E_8$ are in their pure Coulomb phases. To parameterize this branch, we denote by $Y^+_{4i}$ and $Y^+_{5i}$ the Wilson lines along $T^2$ of the $U(1)_{\rm en}$ factor associated to the direction $X^i$, $i\in\{6,7,8,9\}$. As before, the background deformations (\ref{bg}) and (\ref{wl}) are also allowed.  
In the background $(a)$, the massless states $s_0$  charged under each $U(1)_{\rm en}$ are the quaternions of charges $\pm \sqrt{2}$. Their towers of Kaluza-Klein states along~$T^2$ have masses deformed by the Wilson lines $Y^+_{4i},Y^+_{5i}$ and, thanks to Eq. (\ref{Vi}), the untwisted part of the effective potential, $\V^{\N=2\to 0,\rm u}_{\mbox{\scriptsize 1-loop}}$,  yields a contribution 
\be
{1\over 16\pi^5}\,{ M_{\rm s}^4\over \Im T_1}\, E_{(1,0)}(U_1\abs 2,0)\;\sum_{i=6}^9 4\; {1\over 2}\!\sum_{Q_i=\pm\sqrt{2}}\!\left(2\big\abs Y^+_{i}\, Q_{i}\big\abs^2-\rho\big(Y^+_{i}\, Q_{i}\big)^2-\bar\rho\big(\bar Y^+_{i}\, Q_{i} \big)^2\right),
\ee
where   
\be
Y^+_i={U_1 Y^+_{4i}-Y^+_{5i}\over \sqrt{\Im T_1\Im U_1}}\, ,\qquad i\in\{6,\dots,9\}\, . 
\ee 
In the pure Coulomb branch of the gauge theory, the total 1-loop effective potential is therefore
\begin{align}
\label{Vtotc}
\V_{\mbox{\scriptsize 1-loop}}^{\N=2\to 0}= &\; {1\over 16\pi^5}\,{ M_{\rm s}^4\over \Im T_1}\, E_{(1,0)}(U_1\abs 2,0)\Bigg[   \sum_{i=6}^{9} c_{U(1)_{\rm en}}\Big(2\abs Y^+_i\abs^2-\rho(Y^+_i)^2-\bar\rho(\bar Y^+_i)^2\Big)\desp\nonumber  \\
&+c_{E_7} \sum_{\I=10}^{16}\Big(2\abs Y_\I\abs^2-\rho(Y_\I)^2-\bar\rho(\bar Y_\I)^2\Big)+ c_{SU(2)} \Big(2\abs Y_{17}\abs^2-\rho(Y_{17})^2-\bar\rho(\bar Y_{17})^2\Big)\nonumber\tesp\\
&+c_{E_8} \sum_{\I=18}^{25}\Big(2\abs Y_\I\abs^2-\rho(Y_\I)^2-\bar\rho(\bar Y_\I)^2\Big)\Bigg]\!+\cdots+\O\!\left({c^2\Ms^4\over \Im T_1}\, e^{-c\sqrt{\Im T_1}}\right) .
\end{align}

Of course, each Abelian factor of $U(1)_{\rm en}^4$ can be in its own Higgs or Coulomb phase, independently of the others. Thus, there exist a pure Higgs, a pure Coulomb and mixed Coulomb/Higgs branches that realize the spontaneous breaking of the gauge symmetry  $U(1)_{\rm en}^4\to U(1)_{\rm en}^k$, $k\in\{0,\dots, 4\}$. In each case, the 1-loop effective potential takes a form similar to Eqs (\ref{Vtot}) and (\ref{Vtotc}). Some remarks are in order : 

$\bullet$ The pure Higgs branch of the $U(1)_{\rm en}^4$ gauge theory is of real dimension $4\times 4$, parameterized by $Y_{ji}$, $j,i\in\{6,7,8,9\}$. 

$\bullet$ Compactifying for convenience the background $(a)$ down to two spacetime dimensions, the pure Coulomb branch of the $U(1)_{\rm en}^4$ gauge theory is of real dimension $4\times 4$, parameterized by the Wilson lines $Y^+_{ji}$, $j\in\{2,3,4,5\}$, $i\in\{6,7,8,9\}$. 

$\bullet$ Similarly, for the backgrounds $(b)$, the pure Higgs branch of the $SU(2)_{\rm en}\times U(1)_{\rm en}$ gauge theory is of real dimension $4\times 4$, parameterized by $Y_{ji}$, $j,i\in\{6,7,8,9\}$. 

$\bullet$ In the backgrounds $(b)$, the pure Coulomb branch of the $SU(2)_{\rm en}\times U(1)_{\rm en}$ gauge theory describes a  $U(1)_{\rm en}^2$ gauge symmetry with 2  massless neutral quaternions (see the paragraph above Eq.~(\ref{gg}), where the other degrees of freedom have become massive). Compactifying  the model down to two spacetime dimensions, this Coulomb branch is of real dimension $4\times 4$, parameterized by the Wilson lines along $X^{2,3,4,5}$ of the two Abelian vector fields and the 2 quaternions.

The ubiquity of the $4\times 4$ dimension in the above branches is not accidental. In two dimensions, the internal space is 
\be
\label{orb}
T^4\times {T^4\times T^{16}_R\over \Z_2}\, , 
\ee
where the first $T^4$ refers to the directions $X^{2,3,4,5}$, the second one to $X^{6,7,8,9}$ and the last one stands for the right-moving coordinates $X^{10,\dots, 25}$ of the bosonic string. In the parent $\N=4\to 0$ theory, the second $T^4$ yields a gauge theory with gauge group  $G^{(T^4)}$ of rank 4, which in general can be at least $U(1)^4$ (for which $N^\pm=0$) and at most $SO(8)$ (for which $N^\pm =12$). The corresponding degrees of freedom are those of the bosonic parts of $\N=4$ vector multiplets\footnote{We keep counting the supersymmetry generators as in four dimensions.} 
\be
\psi^j_{-1/2}\otimes \bar \O^a\, , \qquad j\in\{2,3,4,5\}\cup \{6,7,8,9\}\, , \; a\in Adj_{G^{(T^4)}}\, , 
\ee 
where $\bar \O^a$ are right-moving operators realizing the affine Kac-Moody algebra of  $G^{(T^4)}$, whose adjoint representation is denoted $Adj_{G^{(T^4)}}$. The deformations of this theory are in one-to-one correspondence with the marginal operators 
\be
\label{pO}
\psi^j_{-1/2}\otimes \bar \O^i\, , \qquad j\in\{2,3,4,5\}\cup \{6,7,8,9\}\, , \; i\in \mbox{Cartan of $G^{(T^4)}$}\, , 
\ee 
where ``Cartan of $G^{(T^4)}$" refers to a  four-dimensional basis of Cartan generators. In the $\N=4\to 0$ theory, all choices of Cartan bases are equivalent. However, from the point of view of the descendent $\N=2\to 0$ model, this is not the case. Without loss of generality,  let us choose a basis whose elements  $\O^i$ are eigenstates of the $\Z_2$ generator\footnote{The $\Z_2$ generator can always be diagonalized in any Cartan subalgebra.}~:

$\bullet$ For instance, assume that all Cartan generators of the basis are odd under $\Z_2$. Then,  the surviving marginal deformations in the $\N=2\to 0$ model are $\psi^j_{-1/2}\otimes \bar \O^i$, $j\in\{6,7,8,9\}$, $i\in \mbox{Cartan of $G^{(T^4)}$}$, and the associated moduli are 4 quaternions. In this case, there are no Wilson line  in the deformed descendant gauge theory, which implies the rank of the gauge group vanishes \ie that the group   is trivial. However, before deformation, the descendent gauge theory describes a gauge symmetry of dimension $N^+$, with $N^-\!+4$ quaternions (see Eq.~(\ref{uspec})). Therefore, the deformations under consideration parameterize the pure Higgs branch of this theory, where the bosonic parts of $N^+$ $\N=2$ vector multiplets combine with those of $N^-=N^+$ charged quaternions  to become massive, thus living us with 4 neutral quaternions.  From Eq. (\ref{z4}), we see that the ``natural" basis of Cartan generators singled out by the right-moving oscillators at level 1 are all odd under the $\Z_2$ generator. The resulting quaternionic coordinates are $Y_{ji}$, $j,i\in\{6,7,8,9\}$. 

$\bullet$ On the contrary, assume that all Cartan generators of the basis are even under $\Z_2$. Then, the surviving marginal deformations in the descendent theory are $\psi^j_{-1/2}\otimes \bar \O^i$, $j\in\{2,3,4,5\}$, $i\in \mbox{Cartan of $G^{(T^4)}$}$, which describe Wilson lines along the untwisted $T^4$. In this case, all moduli sit in the bosonic parts of four $\N=2$ vector multiplets. This shows that the rank of the gauge group in the deformed gauge theory is 4, which is the maximal allowed value, and no condensation of quaternions is allowed. However, before deformation, the descendent gauge theory describes a gauge symmetry of dimension $N^+$, with $N^-\!+4$ quaternions. Therefore, the deformations under consideration parameterize the pure Coulomb branch, where the bosonic parts of $N^+\!-4$ non-Cartan $\N=2$ vector multiplets together with $N^-\!+4$ charged quaternions become massive, thus living us with the bosonic parts of $U(1)_{\rm en}^4$ $\N=2$ vector multiplets. An example of this scenario is provided by the background~$(a)$, since the $N^+=4$ generators even under the $\Z_2$ action  (see Eq. (\ref{z4})) can be used as a Cartan basis of $G^{(T^4)}$. The resulting coordinates are $Y^+_{ji}$, $j\in\{2,3,4,5\}$, $i\in\{6,7,8,9\}$. Note however that it is not always possible to built up a Cartan basis out of generators all even under the orbifold action, as seen with the background $(b)$. 

In fact, for any $\Gamma_{4,4}$ lattice (associated to the second $T^4$ in Eq. (\ref{orb})) that yields a gauge symmetry~$G^{(T^4)}$  in the parent $\N=4\to 0$ theory, all possible choices of Cartan bases fall into 5 equivalence classes, which are characterized by the number $k\in\{0,\cdots, 4\}$ of eigenvalues~$+1$ of the $\Z_2$  action  on the basis. In the  $\N=2\to 0$ $\Z_2$-orbifold model, each non-empty class yields a phase along which  the descendant gauge theory can be deformed. The generic spectrum along each phase amounts to the bosonic parts of $U(1)^k$ $\N=2$ vector multiplets and~$4-k$ neutral quaternions.\footnote{When $k\ge 2$, enhanced gauge symmetries arise when  Wilson lines take equal values.}  In two dimensions, the Coulomb, Higgs or mixed phases of these gauge theories all have real dimension $4\times 4$, which is nothing but~${1\over 2}$ of the number of marginal deformations (\ref{pO}) in the parent $\N=4\to 0$ model. In four dimensions, the real dimension of the branch is $4\times 4-2k$.

Coming back to the four-dimensional case, we introduce dimensions to the scalar fields and conclude from Eq. (\ref{Vtot}) that the Higgs moduli fields $\Ms Y_{ji}$, $j,i\in\{6,7,8,9\}$, of the~$U(1)_{\rm en}^4$ gauge theory acquire masses proportional to $\m$ at 1-loop.  Since $\abs\rho\abs<1$, the quadratic form  $2\abs Y\abs^2-\rho Y^2-\bar\rho\bar Y^2$ has two positive eigenvalues.  As a result, thanks to Eq.~(\ref{Vtotc}), the Coulomb branch moduli of the $U(1)_{\rm en}^4\times SU(2)\times E_8$ gauge symmetry are also massive, while those of $E_7$ are tachyonic. Only $T_1,U_1$, which parameterize the Coulomb branch of $U(1)^2$, and in particular their combination $\m$, remain massless at $\mbox{1-loop}$.\footnote{To be precise, contrary to the other moduli of the background $(a)$, they admit tadpoles (as well as mass terms), which are however exponentially suppressed when $\m$ is low. This is due to the fact that the~$U(1)^2$ gauge symmetry arising from the large $T^2$ lattice is not enhanced \cite{GV}.}  From a dynamical point of view, the massive scalars are attracted to the origin of their respective branches \cite{attractor,LP,Patil:2004zp}, $\Ms Y_{ji}=\Ms Y_{i}=\Ms Y_{17}=\dots =\Ms Y_{25}=0$, while the tachyonic ones  $\Ms Y_{10}, \dots, \Ms Y_{16}$ develop expectation values that break $E_7$ to some subgroup of rank~7. Thus,  the background ($a$) admits instabilities at the quantum level, which imply the potential to become negative. 

The dependance on the quadratic charges of the positive or negative squared masses of the moduli can be naturally written in terms of $\beta$-function coefficients of the undeformed background. Keeping things general in this paragraph, we consider any $\Z_2$-orbifold no-scale model, where the $\N=2\to 0$ Scherk-Schwarz breaking is implemented along $T^2$ \via the phase $SS'$. For any gauge group factor $\H$ of the undeformed $\N=2\to 0$ theory ($U(1)_{\rm en}$, $E_7$, $SU(2)$ or $E_8$ in our example), there is a gauge symmetry $\G$ in the parent $\N=4\to 0$ model ($SU(2)_{\rm en}$ or $E_8$ in our example) such that the dressing coefficient of the mass terms of the moduli related to $\H$ is  
\be
\label{cb}
c_\H={8\over 2}\, C_{Adj_\G}-4\, C_{{\cal R}^{\rm t}_{\H}} = -3\left( {b_\G\over 2}+b_\H^{\rm t}\right).
\ee
In this relation, $Adj_\G$ is the adjoint representation of $\G$ realized by the bosonic parts of $\N=4$ vector multiplets, with corresponding $\beta$-function coefficient $b_\G$. ${\cal R}^{\rm t}_\H$ is the representation of~${\cal H}$ realized by the fermionic parts of the twisted hypermultiplets, whose contribution to the $\mbox{$\beta$-function}$ coefficient of $\H$ is $b_\H^{\rm t}$. To derive Eq. (\ref{cb}), we use the fact that massless degrees of freedom in a representation ${\cal R}_{\cal K}$ of any gauge group ${\cal K}$ contribute to the $\beta$-function coefficient of $\cal K$ as  
\be
 \label{coeff}
b_{\cal K}^{\mbox{\scriptsize gauge boson}} =-{11\over 3} \, C({{\cal R}_{\cal K}})\, , \quad   b_{\cal K}^{\mbox{\scriptsize real scalar}} ={1\over 6} \,  C({{\cal R}_{\cal K}})\, , \quad b_{\cal K}^{\mbox{\scriptsize Majorana fermion}} = {2\over 3}\, C({{\cal R}_{\cal K}})\, . 
\ee

In Ref. \cite{SNSM}, it is shown that in the $\N=4\to 0$ super no-scale models, the Wilson lines associated to the asymptotically free gauge theories are stable. In this case, the non-Abelian gauge symmetries are expected to confine at low energy. On the contrary, the non-asymptotically free gauge theories yield Wilson line instabilities, which should survive in the infrared. Turning back to the $\N=2\to 0$ super no-scale models, we find in the background~$(a)$ the  $\beta$-function coefficients  
\be
b_{U(1)_{\rm en}}={4\over 3}\, , \qquad b_{E_7}=12\, , \qquad b_{SU(2)}={100\over 3}\, , \qquad b_{E_8}=-100\, , 
\ee
while those of $U(1)^2_{\rm grav,ten}\times U(1)^2$  are vanishing. Thus, at low energy, the $E_8$ gauge symmetry should confine, while our description of the enhanced $U(1)_{\rm en}^4$ and $SU(2)$ gauge symmetries, together with that of the spontaneous breaking of $E_7$ to a rank 7 subgroup, are expected to be valid.
 
\vspace{.2cm}
{\large \em \noindent Mixed Coulomb/Higgs phases of $E_7\times SU(2)$}

\noindent  In the background $(a)$, the phases of the gauge symmetry (\ref{gga}) that remain to be described are those where the  $E_7\times SU(2)$ group is spontaneously broken to some subgroup of rank~$\mbox{$r< 8$}$. This happens when degrees of freedom in the $[56]_{E_7}\otimes [2]_{SU(2)}$  bifundamental representation condense. However, the $E_7\times SU(2)$ gauge theory realized in the $\N=2\to 0$ $\Z_2$-orbifold model involves untwisted states only and is therefore obtained by truncation of the parent $\N=4\to 0$ model. This means that its phase structure is expected to be similar to that presented for the $U(1)_{\rm en}^4$ gauge symmetry coupled to charged quaternions. 

In the parent $\N=4\to 0$ theory, the right-moving $T_R^{16}$ in  Eq. (\ref{orb}) yields the $E_8\times E_8$ gauge symmetry. In two dimensions, the degrees of freedom, and among them the marginal deformations, are in one-to-one correspondence with the operators 
\begin{align}
\label{pE}
&\psi^j_{-1/2}\otimes \bar \O^a\, , \qquad j\in\{2,3,4,5\}\cup \{6,7,8,9\}\, , \; a\in Adj_{E_8\times E_8}\, , \nonumber \\
&\psi^j_{-1/2}\otimes \bar \O^i\, , \,\qquad j\in\{2,3,4,5\}\cup \{6,7,8,9\}\, , \,\; i\in \mbox{Cartan of $E_8\times E_8$}\, . 
\end{align}
Since the orbifold action is trivial on the second  $E_8$ factor, its generators are all invariant under the $\Z_2$ twist. Thus, any choice of Cartan basis of the second $E_8$ contains even generators only, which shows that the basis falls into a unique equivalence class. In the descendent $\N=2\to 0$ model, the surviving marginal deformations of the second $E_8$ gauge theory are therefore  of the form $\psi^j_{-1/2}\otimes \bar \O^i$, $j\in\{2,3,4,5\}$, $i\in \mbox{Cartan of second $E_8$}$. They are associated to Wilson lines along the untwisted $T^4$ in Eq. (\ref{orb}), and  parameterize the pure Coulomb phase of the second $E_8$. 

On the contrary, the $\Z_2$ action is non-trivial on the characters of the first $E_8$, and thus on the associated right-moving operators $\bar \O^a$, $a\in Adj_{E_8}\equiv [248]_{E_8}$. Actually, 
since $\mbox{$E_7\times SU(2)\subset E_8$}$, we can decompose 
\be
[248]_{E_8}=[133]_{E_7}\oplus [3]_{SU(2)}\oplus [56]_{E_7}\otimes [2]_{SU(2)}
\ee
and see from the spectrum in Eq. (\ref{uspec}) that $\bar \O^a$, $a\in [133]_{E_7}\oplus[3]_{SU(2)}$, are even under the $\Z_2$~generator, while $\bar \O^a$, $a\in [56]_{E_7}\otimes[2]_{SU(2)}$ are odd. Therefore, any choice of $E_8$ Cartan basis  in the parent theory, which contains $8-r$, $r\in \{0,\dots,8\}$,  generators in the $[56]_{E_7}\otimes[2]_{SU(2)}$ yields in the descendent model $8-r$ quaternionic moduli and $r$ Wilson line deformations along the untwisted $T^4$. Therefore the $E_7\times SU(2)$ gauge theory is in a mixed phase, where the generic spectrum amounts to the bosonic parts of $U(1)_{\rm en}^r$ $\N=2$ vector multiplets and $8-r$ neutral quaternions. From the two-dimensional viewpoint,  the Coulomb, Higgs or mixed phases of the $E_7\times SU(2)$ gauge theory all have real dimension $4\times 8$, which is nothing but ${1\over 2}$ of the number of marginal deformations in the parent $\N=4\to 0$ model.  In four dimensions, the real dimension of the branch is $4\times 8-2r$.


\section{Descendent \bm $\N=1\to 0$ super no-scale  models}
\label{1->0}

In this section, we would like to justify that the $\N=2\to 0$ super no-scale models realized as $T^2\times T^4/\Z_2$ compactifications with stringy Scherk-Schwarz mechanism along $T^2$  yield descendent $\N=1\to 0$ super no-scale models, once a second  orbifold twist is implemented. However, we argue that the resulting 1-loop effective potential requires further study. 

Starting with any $\N=2$ heterotic model compactified on $T^2\times (T^2\times T^2)/\Z_2$, where the $\Z_2$ twist generator is denoted $\boldsymbol{G}$, one obtains  an $\N=2\to 0$ no-scale model by implementing a stringy Scherk-Schwarz mechanism along the first $T^2$. Contrary to Sect. \ref{2->0}, we do not suppose that all massless bosons in the twisted  sector acquire a tree level mass of order~$\m$, so that twisted moduli may exist. We consider the descendent  $\N=1\to 0$ no-scale model obtained by implementing a $\Z_2'$ orbifold twist of generator $\boldsymbol{G}'$, which acts on the first and third~$T^2$'s. In this case, the 1-loop effective potential of the $\N=1\to 0$ model can be written as
\be
\Vone^{\N=1\to 0} =-{M_{s}^4\over (2\pi)^4}\int_\F {d^2\tau\over 2\tau_2^2}\left[ \underset{\rm untswisted'}{\Str}\!\!\!\!{1+\boldsymbol{G'}\over 2} q^{{1\over 4}M_L^2/\Ms^2}\bar q^{{1\over 4}M_R^2/\Ms^2} +\! \underset{\rm twisted'}{\Str}\!{1+\boldsymbol{G'}\over 2} q^{{1\over 4}M_L^2/\Ms^2}\bar q^{{1\over 4}M_R^2/\Ms^2}  \right]\!,
\ee
where $M_L,M_R$ are the left- and right-moving masses. In this expression, ``untwisted$'$" denotes the spectrum of the parent $\N=2\to 0$ model, while ``twisted$'$" refers to the twisted spectrum, with respect to $\Z_2'$. Since $\boldsymbol{G'}$ twists the first $T^2$, the twisted$'$ states invariant under $\boldsymbol{G'}$ have vanishing momenta and winding numbers along the directions $X^{4,5}$. Therefore, their tree-level masses are independent of $\m$ and are supersymmetric. In other words, the bosons/fermion degeneracy in this sector is not lifted classically.\footnote{The spontaneous breaking of supersymmetry is mediated to the twisted$'$ sector by quantum interactions with the non-supersymmetric untwisted$'$ sector.}  This shows that the ${\Str}$ over the twisted$'$ sector with the $(1+\boldsymbol{G'})/2$ projector inserted is vanishing. Since the corresponding conformal blocks form an $SL(2,\Z)$ modular orbit with the $\Str$ over the untwisted$'$ sector with $\boldsymbol{G'}$ inserted, this second $\Str$ is also vanishing. Therefore, 
\be
\label{pot}
\Vone^{\N=1\to 0} =-{M_{s}^4\over (2\pi)^4}\, \,{1\over 2} \int_\F {d^2\tau\over 2\tau_2^2}\, \underset{\rm untswisted'}{\Str}\!\!\!\! q^{{1\over 4}M_L^2/\Ms^2}\bar q^{{1\over 4}M_R^2/\Ms^2}  ,
\ee
which is an exact identity, no matter the scale $\m$ is, compared to $\Ms$. It is valid for arbitrary moduli deformations of the parent $\N=2\to 0$ model that survive the $(1+\boldsymbol{G'})/2$ projection.\footnote{This includes the complex structure deformations of $(T^2\times T^2)/\Z_2$ into a smooth K3 surface. When they are tuned on,  the $\Z_2'$ action must be suitably defined at the level of the CFT, but the key point is that it still acts as a geometrical twist on the $T^2$ along which the stringy Scherk-Schwarz mechanism is implemented. In this case, ``untwisted$'$"  denotes again the spectrum of the parent $\N=2\to 0$ model. In the class of $\N=2\to 0$ $\Z_2$-orbifold models defined in Sects \ref{2->0}, where all twisted moduli are lifted classically, such complex structure deformations  do not exist.} An important consequence of this equation is  that if the parent $\N=2\to 0$ theory sits at a point in moduli space where it develops a super no-scale structure, then the descendent $\N=1\to 0$ model is also super no-scale. 

Moreover, Eq. (\ref{pot})  suggests we may write  
\be
\label{210}
\Vone^{\N=1\to 0}= {1\over 2}\, \Vone^{\N=2\to 0}\, .
\ee
However, it is important to stress that this is only possible if new moduli of the $\N=1\to 0$ theory are not switched on. The latter may arise from the twisted$'$ sector of the theory, whose tree level mass spectrum is not affected by the Scherk-Schwarz breaking. To be more specific, the group elements $\boldsymbol{G'}$ and $\boldsymbol{GG'}$ may admit fixed points (copies of the second and third $T^2$'s, respectively) and thus introduce new massless $\N=1$ chiral supermultiplets. By switching on vacuum expectations values to their bosons, the gauge symmetry of the $\N=1\to 0$ model (which arises from the untwisted$'$ sector) enters Higgs branches, which are parameterized by moduli having  no counterpart in the parent $\N=2\to 0$ theory. Note that these moduli are complex structure deformations of the first and third $T^2$'s modded out by $\Z_2'$, or the first and second $T^2$'s modded out by the diagonal subgroup of $\Z_2\times \Z_2'$.  Since the masses $M_L,M_R$ of the initial $\N=2\to 0$ untwisted$'$ sector depend on these moduli, Eq.~(\ref{pot}) is a quantum potential for these deformations. As a result, the latter acquire positive or negative squared masses at 1-loop, which yield additional conditions for the background to be stable.
To evaluate these quantum masses, one may again apply Eq. (\ref{Vi}), dressed with an overall factor ${1\over 2}$, but with the masses $M_L$ now depending of these new moduli. We mention that the dependance of the $M_L$'s on these deformations can be determined in the classical effective $\N=1$ gauged  supergravity at low energy, by following the method applied in Ref. \cite{LP} in a similar problem.

\vspace{.2cm}
{\large \em \noindent A class of $\N=1\to 0$ super no-scale models}

\noindent  To proceed, we focus on the $\N=2\to 0$ super no-scale $\Z_2$-orbifold models presented in Sect~\ref{2->0}, where all moduli arise in the untwisted sector, and construct descendent $\N=1\to 0$ super no-scale models. Compactifying down to two dimensions, the internal space is
\be
\label{orb2}
T^2\times {T^2\times T^2\times T^2\times T^{16}_R\over \Z_2\times \Z_2'}\, , 
\ee
where the first $T^2$ refers to the directions $X^{2,3}$, the second one to $X^{4,5}$, the third one to $X^{6,7}$ and the last one to $X^{8,9}$. As before, the stringy Scherk-Schwarz mechanism is implemented along the second $T^2$. At the $\N=4\to 0$ level \ie without implementation of the $\Z_2\times \Z_2'$ action, the right-moving coordinates of the last three $T^2$'s as well as $T^{16}_R$ generate a gauge symmetry $\G$ or rank 22. Choosing a Cartan subalgebra,  and a basis for it which diagonalizes $\boldsymbol{G}$ and $\boldsymbol{G'}$, we can impose the  projector 
\be
\left({1+\boldsymbol{G'}\over 2}\right)\left({1+\boldsymbol{G}\over 2}\right)
\ee
on the set of marginal operators of the parent $\N=4\to 0$ model, and find those which survive in the descendent $\N=1\to 0$ theory, namely
\be
\label{proj}
\psi^j\otimes \bar\O^i\, , \quad i\in \mbox{Cartan of $\G$} \, , \; j\in\left\{
\begin{array}{l}
\{2,3\} \quad \mbox{if} \;\; \boldsymbol{G} \O^i=+\O^i \;\; \and\;\;  \boldsymbol{G'} \O^i=+\O^i\, , \\
\{4,5\} \quad \mbox{if} \;\;\boldsymbol{G} \O^i=+\O^i \;\;\and\;\; \boldsymbol{G'} \O^i=-\O^i\, ,\\
\{6,7\} \quad \mbox{if} \;\;\boldsymbol{G} \O^i=-\O^i \;\;\and\;\; \boldsymbol{G'} \O^i=+\O^i\, , \\
\{8,9\} \quad \mbox{if} \;\;\boldsymbol{G} \O^i=-\O^i \;\;\and\;\; \boldsymbol{G'} \O^i=-\O^i\, .
\end{array}\right. 
\ee
We see that the Cartan generators even under $\boldsymbol{G}$ and $\boldsymbol{G'}$ are associated to vector boson Wilson lines along the first $T^2$, while all other Cartan generators yield complex scalar deformations. The Wilson lines parameterize the Coulomb phase associated to the $\bar \O^i$'s that remain Cartan generators of the gauge symmetry in the $\N=1\to 0$ model, while the complex scalars span the Higgs phase associated to the $\bar \O^i$'s that are no more generators of the descendent gauge symmetry. The real dimension of each phase is $2\times 22$, which is nothing but ${1\over 4}$ of the number of marginal deformations in the parent $\N=4\to 0$ model.  

Turning back to the four-dimensional case, the deformations along $X^{2,3}$ become transverse gauge degrees of freedom and the Coulomb phases zero-dimensional (!) There is no pure Coulomb branch, no mixed Coulomb/Higgs branch, and the only phase that exists is the pure Higgs one. Actually, all Cartan subalgebras of $\G$ that share a common pattern of~$(\pm1,\pm1)$ eigenvalues with respect to $\boldsymbol{G},\boldsymbol{G'}$ yield, in the descendent $\N=1\to 0$ theory,  a gauge symmetry of rank $r$, $r\in\{0,\dots,22\}$, where $r$ counts the number of~${(+1,+1)}$ pairs of eigenvalues. However, this does not mean that for different ranks $r<r'$, the $\N=1\to 0$ background sits in different branches of its gauge theory. 
Instead, when the rank is $r'$, the model still sits on the locus where the gauge symmetry is generically of rank $r$, but on a real codimension $2(r'-r)$ submanifold, where the  gauge symmetry is enhanced to onother one of rank~$r'$.

\vspace{.2cm}
{\large \em \noindent Descendants of the backgrounds $(a)$ and $(b)$}

\noindent  In order to illustrate the above analysis, we construct the $\N=1\to 0$ super no-scale models obtained by implementing the $\Z_2'$ action on the $\N=2\to 0$ super no-scale $\mbox{$\Z_2$-orbifold}$  backgrounds $(a)$ and $(b)$. 
Proceeding as in Sect. \ref{2->0}, the 1-loop partition function can be expressed in terms of $SO(10)\times SO(2)\times SO(2)\times SO(2)$ affine characters. In the untwisted sector (with respect to $\Z_2$ and $\Z_2'$) of the descendant $\N=1\to 0$  model, all fermions of the initially massless $\N=1$ supermultiplets acquire a tree level mass from the stringy Scherk-Schwarz mechanism implemented with the sign $SS'$ in the partition function. In the backgrounds $(a)$ and $(b)$, the internal $T^4$ associated to the directions $X^{6,7,8,9}$ being factorized as $T^2\times T^2$, the gauge group $G^{(T^4)}$ of the parent $\N=4\to 0$ model can be written as~$G^{(T^2)}_{(2)}\times G^{(T^2)}_{(3)}$. We define 
\be
N^\pm_{(2)}={\dim G^{(T^2)}_{(2)}-2\over 2}\, , \qquad N^\pm_{(3)}={\dim G^{(T^2)}_{(3)}-2\over 2}\, ,
\ee
where $N^\pm_{(2)}$ are the numbers of non-Cartan generators of $G^{(T^2)}_{(2)}$ that are even or odd under $\boldsymbol{G}$ and~$\boldsymbol{GG'}$, while $N^\pm_{(3)}$ are similarly the numbers of non-Cartan generators of $G^{(T^2)}_{(3)}$ that are even or odd under $\boldsymbol{G}$ and $\boldsymbol{G'}$. In terms of these notations, the representations of the untwisted massless states of the $\N=1\to 0$ undeformed backgrounds are
\begin{align}
\label{uspec''}
\mbox{Bosons in\;\;\;\;\;}&\; [2]_{\psi^{2,3}_{-1/2}}\otimes \Big([2]_{\bar X^{2,3}_{-1}}\oplus [N^{+}_{(2)}]\oplus [N^{+}_{(3)}]\oplus [78]_{E_6}\oplus [1]_{U(1)}\oplus [1]_{U(1)'}\oplus [248]_{E_8}\Big)\nonumber\\
\oplus&\; [2]_{\psi^{4,5}_{-1/2}}\otimes \Big([2]_{\bar X^{4,5}_{-1}}\phantom{\oplus\; \,  [N^-_{(2)}]}\oplus [27]^{{s\over 2},{s'\over 2}}_{E_6}\oplus [\overline{27}]^{-{s\over 2},-{s'\over 2}}_{E_6}\oplus [1]^{-{s\over 2},{3\over 2}s'}_{E_6}\oplus [1]^{{s\over 2},-{3\over 2}s'}_{E_6}\Big)\nonumber \\
\oplus&\; [2]_{\psi^{6,7}_{-1/2}}\otimes \Big([2]_{\bar X^{6,7}_{-1}}\oplus [N^-_{(2)}]\oplus [27]^{-{s\over 2},{s'\over 2}}_{E_6}\oplus [\overline{27}]^{{s\over 2},-{s'\over 2}}_{E_6}\oplus [1]^{{s\over 2},{3\over 2}s'}_{E_6}\oplus [1]^{-{s\over 2},-{3\over 2}s'}_{E_6}\Big)\nonumber \\
\oplus&\; [2]_{\psi^{8,9}_{-1/2}}\otimes \Big([2]_{\bar X^{8,9}_{-1}}\oplus [N^-_{(3)}]\oplus [27]^{0,-s'}_{E_6}\oplus [\overline{27}]^{0,s'}_{E_6}\oplus [1]^{s,0}_{E_6}\oplus [1]^{-s,0}_{E_6}\Big).
\end{align}
This spectrum amounts to the bosonic parts of $\N=1$ supermultiplets : 1 gravity multiplet (graviton), 1 linear multiplet (antisymmetric tensor, dilaton), 1 vector multiplet (gauge boson) in the adjoint representation of  $E_6\times U(1)\times U(1)'\times E_8$, 3 chiral multiplets (complex scalar) in the  $[27]_{E_6}\oplus[\overline{27}]_{E_6}$ and 3 chiral multiplets in the $[1]_{E_6}\oplus[1]_{E_6}$, whose charges under~$U(1)\times U(1)'$ are indicated in upper indices (with $s=\sqrt{2}$, $s'=\sqrt{6}/3$), and 2 neutral chiral multiplets associated to the moduli $T_1,U_1$. The remaining supermultiplets depend on $G^{(T^2)}_{(2)}$ and $G^{(T^2)}_{(3)}$. They are identical to those found in the parent $\N=2\to 0$ background, up to the replacement of the $\N=2$ vector multiplets and hypermultiplets with  $\N=1$ vector multiplets and chiral multiplets. For instance, for the background ($a$), we have $G^{(T^2)}_{(2)}=G^{(T^2)}_{(3)}=SU(2)^2_{\rm en}$, so that $N^+_{(2)}=N^+_{(3)}=2$, which give $U(1)_{\rm en}^4$  Abelian vector multiplets. We also have $N^-_{(2)}+2=N^-_{(3)}+2=4$, which yield for each of these enhanced $U(1)_{\rm en}$ factors 1~pair of chiral multiplets of charges $\pm \sqrt{2}$. Note that  at the exact $\N=1$ level, \ie without implementation of the stringy Scherk-Schwarz breaking, all untwisted chiral multiplets would be in  non-chiral representations of the gauge group. Some remarks are in order : 

$\bullet$ Since the first $T^2$ is very large and does not sit at any enhanced symmetry point, the rank 2 gauge group it may describe is Higgsed. The corresponding phase is parameterized by $T_1,U_1$ \ie $2\times 2$ real moduli. 

$\bullet$ The Higgs phase of the $U(1)_{\rm en}^4$ gauge symmetry of rank 4 is parameterized by  $Y_{ji}$, $j,i\in\{6,7\}$ and $\mbox{$j,i\in\{8,9\}$}$ \ie $2\times 4$ real moduli. 

$\bullet$ The 248 generators of the $E_8$ gauge symmetry are even under $\boldsymbol{G}$ and $\boldsymbol{G'}$. Thus,  no choice of Cartan subalgebra can yield complex scalars deformations, and the $E_8$ gauge symmetry cannot be Higgsed. As a result, the dimension of the whole Higgs branch of the model cannot exceed $2\times 14$. Moreover, the rank of the whole gauge group is at least 8 everywhere in the branch. 

$\bullet$ The $E_6\times U(1)\times U(1)'$ gauge symmetry can be Higgsed to subgroups of rank $r<8$ by the charged complex scalars. Knowing if $r$ admits a lower bound $r_m>0$ requires more study. However, if this is the case, the upper bound on the real dimension of the Higgs phase of the model becomes $2\times (14-r_m)$, while the minimal rank everywhere in moduli space becomes $8+r_m$.

Moreover, the generator $\boldsymbol{G}$ fixes geometrically 16 copies of the first $T^2$. In the associated twisted sector, due to the implementation of the Scherk-Schwarz mechanism with sign insertion $SS'$ in the partition function, all bosons of the initially massless $\N=1$ supermultiplets acquire a tree level mass of order $\m$. The massless states are Weyl fermions of initially $\N=1$ chiral multiplets, where the first components are in the representations 
\be
\label{p1}
16 [27]^{-{s\over 4},-{s'\over 4}}_{E_6}\oplus 16 [1]^{{3\over 4}s,{3\over 4}s'}_{E_6}\oplus 32 [1]^{{s\over 4},-{3\over 4}s'}_{E_6}\oplus 32 [1]^{-{s\over 4},{3\over 4}s'}_{E_6}
\ee
and the second ones in the conjugate representations. Those with degeneracy 16 yield chiral families, while those with degeneracy 32 are non-chiral. 

In the same spirit, the generators $\boldsymbol{G'}$ and $\boldsymbol{GG'}$ fix 16 copies of the second and third $T^2$'s, respectively. The associated twisted sectors are similar to that fixed by $\boldsymbol{G}$, up to the fact that the tree level masses are not affected by the stringy Scherk-Schwarz mechanism. Together, they are nothing but the  twisted$'$ sector of the $\N=1\to 0$ model and, at the massless level, contain full $\N=1$ chiral multiplets in representations of $E_6$ as in Eq.~(\ref{p1}), with similar $U(1)\times U(1)'$ charges. 

Assuming the moduli of the twisted$'$ sector are not switched on, we can apply Eq. (\ref{210}) to derive the 1-loop effective potential of the $\N=1\to 0$ super no-scale model that descends from the background $(a)$. We have computed the potential of the parent $\N=2\to 0$ theory when the latter is allowed to be deformed along the branch where $U(1)_{\rm en}^4$ is Higgsed and the $U(1)^2\times E_7\times SU(2)\times E_8$ gauge symmetry is in its $U(1)^{18}$ Coulomb phase. In Eq. (\ref{Vtot}), the moduli that survive the $\Z_2'$ projection are the components of the metric and antisymmetric tensor that respect the $T^2\times T^2\times T^2$ factorization, namely $T_1,U_1$, as well as  $Y_{ji}$, $j,i\in\{6,7\}$ and $j,i\in\{8,9\}$. In particular, as seen in Eq. (\ref{proj}), no Wilson line $Y_\I$, $\I\in\{10,\dots,25\}$  survive. Thus,  we obtain
\be
\label{Vtot1}
\V_{\mbox{\scriptsize 1-loop}}^{\N=1\to 0}= {1\over 16\pi^5}\,{ M_{\rm s}^4\over \Im T_1}\, E_{(1,0)}(U_1\abs 2,0)\Bigg[ 4  \sum_{i,j=6}^7(Y_{ji})^2+ 4  \sum_{i,j=8}^9(Y_{ji})^2\Bigg]\!\!+\cdots+\!\O\!\left({c^2\Ms^4\over \Im T_1}\, e^{-c\sqrt{\Im T_1}}\right)\!.
\ee
In this expression,  the fact that the  complex scalar deformations that parameterize the Higgs phase of $E_6\times U(1)\times U(1)'$ do not appear means  that these moduli are simply set to zero \ie that the descendent model sits at the origin of the Higgs phase of the $E_6\times U(1)\times U(1)'$ gauge symmetry. On the contrary, the complex scalars that span the Higgs phase of the rank 2 group associated to the large $T^2$, as well as the Higgs phase of $U(1)^4_{\rm en}$, are switched on. Thus, the potential (\ref{Vtot1}) is that obtained when the gauge group of the $\N=1\to 0$ model is  enhanced to $E_6\times U(1)\times U(1)'\times E_8$.

Alternatively, we could have allowed the initial parent background $(a)$ to be deformed along its pure Coulomb phase, \ie when $U(1)^2\times U(1)_{\rm en}^4\times E_7\times SU(2)\times E_8\to U(1)^{22}$. In~Eq.~(\ref{Vtotc}), the only moduli that survive the $\Z_2'$ projection are $T_1,U_1$, whose expectation values Higgs    the rank 2 group associated to the large $T^2$.  The effective potential of the descendent model takes therefore  the apparently trivial form
\be
\label{Vtotc1}
\V_{\mbox{\scriptsize 1-loop}}^{\N=1\to 0}= \O\!\left({c^2\Ms^4\over \Im T_1}\, e^{-c\sqrt{\Im T_1}}\right).
\ee
However, since none of the  complex scalar deformations along the Higgs phase of $U(1)_{\rm en}^4\times E_6\times U(1)\times U(1)'$ appears, we conclude that the descendent model sits at the origin of their  Higgs phase and that the gauge symmetry is  enhanced to $U(1)_{\rm en}^4\times E_6\times U(1)\times U(1)'\times E_8$. Therefore, Eq. (\ref{Vtotc1}) does not contain any information that is not already encoded in Eq.~(\ref{Vtot1}). Nonetheless, it is interesting to note that technically, the two expressions are obtained from different choices of Cartan subalgebras of  $\G=U(1)^2\times SU(2)_{\rm en}^4\times E_8\times E_8$ in the $\N=4\to 0$ initial theory, which fall into distinct equivalences classes. 

Actually, the branch of the background $(a)$ that yields the maximum information on the moduli masses of the descendent $\N=1\to 0$ super no-scale model is that where the rank of the gauge symmetry group is minimal. It is obtained when $U(1)_{\rm en}^4$ is totally Higgsed and $E_7\times SU(2)$ is in its Higgs branch of maximal dimension. If we had computed $\V_{\mbox{\scriptsize 1-loop}}^{\N=2\to 0}$  in this phase, we may have found instabilities that induce the breaking of~$E_7$  $\mbox{(or $E_7\times SU(2))$}$ to subgroups of lower ranks. Using this expression of $\V_{\mbox{\scriptsize 1-loop}}^{\N=2\to 0}$ in Eq. (\ref{210}), we may have found instabilities in the $\N=1\to 0$ model, responsible for the (partial) breaking of $\mbox{$E_6\times U(1)\times U(1)'$}$ to subgroups of lower ranks. 


\section{Conclusion}
\label{cl}

The super no-scale models \cite{ADM,planck2015,SNSM,FR}, which by definition have exponentially suppressed effective potential at 1-loop for low supersymmetry breaking scale $\m$, may help to build quantum theories consistent with flat space, as well as to cancel the dilaton tadpole. However, for this to have any chance to work, the backgrounds must be stable at the quantum level. The question of the moduli stability in the $\N=4\to 0$ super no-scale models was addressed in Ref. \cite{SNSM} and the purpose of the present work is to initiate the analysis of the $\N=2\to 0$ case. 

The particular class of models we focus on are heterotic $\Z_2$-orbifolds on $T^2\times T^4/\Z_2$, where the $\N=2\to0$ spontaneous breaking of supersymmetry is implemented \via a stringy Scherk-Schwarz mechanism \cite{SSstring, Kounnas-Rostand} along $T^2$.  We show that a specific implementation of this mechanism induces a mass of order $\m$ to all initially massless bosonic (fermionic) degrees of freedom arising in the twisted (untwisted) sector. Thus, the super no-scale condition, which amounts to canceling the would-be dominant $\m^4$ contribution to the effective potential, is fulfilled by adjusting the number  of massless untwisted bosons to match the number of massless twisted fermions. An obvious but nevertheless fundamental consequence of this choice of supersymmetry breaking  is that all twisted moduli are lifted classically. Moreover, the models do not suffer from classical Hagedorn-like instabilities \cite{Hage}, whatever high $\m$ may be. Actually, the classical tachyons arising  in their parent $\N=4\to 0$ theories (without $\Z_2$ action) when $\m=\O(\Ms)$  are projected out, even if other small marginal deformations are turned on. 

Because in this class of models no twisted deformation exists, the internal $T^4/\Z_2$ cannot be deformed into a smooth $K3$. This implies that their classical vacuum structure is encoded in that of the parent theories. The latter are $\N=4\to 0$ models, where all fermionic degrees of freedom are  massive. In these parent theories, to any  choice of  Cartan generators of the gauge symmetry realized by the bosonic side of the heterotic string, corresponds a set of marginal deformations of the two-demensional worldsheet conformal field theory. All choices of Cartan subalgebras being  equivalent, the vacuum structure is that of a unique Coulomb branch of an $\N=4$ supergravity. In a descendent $\Z_2$-orbifold model, the choices of Cartan subalegras in the parent theory fall into different equivalence classes characterised by their patterns of $\pm1$ eigenvalues with respect to the $\Z_2$ generator. The vacuum structure of the $\N=2\to 0$ models that emerges is that of various mixed Coulomb/Higgs branches, which intersect in moduli space along loci of enhanced gauge symmetry.    

In general, the super no-scale structure emerges precisely at such points of extended gauge symmetry. Thus, the representative points in moduli space of these backgrounds are extrema of the 1-loop effective potential \cite{GV}. When these extrema are saddle or maxima, the $\N=2\to 0$ super no-scale backgrounds are destabilized into either a Coulomb, Higgs or mixed Coulomb/Higgs branch.  The string computation of the 1-loop effective potential  being based on on-shell data at tree level, the result depends on the  branch along which one supposes the theory may be deformed. In a representative example of the class of $\N=2\to 0$ super no-scale models described above, we have evaluated the 1-loop effective potential in the pure Coulomb phase as well as in  mixed Coulomb/Higgs branches. It is enough to derive explicit expressions at quadratic order in moduli fields to conclude on  eventual destabilizations in classically marginal directions that become tachyonic at 1-loop.

We also show that for any implementation of the Scherk-Schwarz mechanism along $T^2$, the $\N=2\to 0$ super no-scale $\Z_2$-orbifold models yield $\N=1\to 0$ super no-scale backgrounds, by implementing  a second $\Z_2$ orbifold action. The models are naturally chiral. Classically, the twisted sector of the second $\Z_2$ remains $\N=1$ supersymmetric and contains new moduli fields. In the present work, the study of the vacuum structure is only partial, in the sense that these new marginal deformations are not switched on. Considering the above-described stringy Scherk-Schwarz breaking of supersymmetry, the orbifold structure is therefore not deformed into a ``non-supersymmetric version of smooth Calabi-Yau compactification". Under these conditions, the pattern of $(\pm 1,\pm 1)$ eigenvalues with respect to the $\Z_2\times \Z_2$ twists can again be found, for any  choice of Cartan subalgebra in the underlying ``grandparent" $\N=4\to0$ model. The resulting structure of classical vacua in the descendent $\N=1\to 0$ theory is that of a unique Higgs branch, inside of which loci where gauge symmetries coupled to charged  complex scalars are restored,  when the expectation values of the latter vanish. Note that at a generic point of the Higgs branch, all gauge group factors with no charged complex scalars remain obviously unbroken, but may confine in the infrared.  

The models we consider in this work are not studied out of the super no-scale regime,~\ie when $\m$ is large enough for the 1-loop effective potential not to be exponentially suppressed. At high supersymmetry breaking scale, at early times in a cosmological scenario, the potential may be positive and drive dynamically the model into the super no-scale regime \cite{SNSM,FR,Aaronson:2016kjm}, or be negative and admit an AdS  vacuum \cite{Angelantonj:2006ut,Aaronson:2016kjm}, or induce large marginal deformations of moduli other than $\m$ and let the model develop a severe Hagedorn-like instability.    


\section*{Acknowledgement}
 
We are grateful to  S. Abel, C. Angelantonj, I. Florakis and  J. Rizos  for fruitful discussions. 
The work of C.K. is partially supported by his  Gay Lussac-Humboldt Research Award 2014, in the  Ludwig Maximilians University and Max-Planck-Institute for Physics. H.P. would like to thank the Laboratoire de Physique Th\'eorique of Ecole Normale Sup\'erieure and the C.E.R.N. Theoretical Physics Department for hospitality. 


\section*{Appendix}
\renewcommand{\theequation}{A.\arabic{equation}}
\renewcommand{\thesection}{A}
\setcounter{equation}{0}
\label{app}

The 1-loop partition function associated to a $d$-dimensional torus can be factorized into a~$\Gamma_{d,d}$ contribution of the lattice of momenta and winding numbers, and a part arising from the left- and right- moving bosonic oscillators. In this Appendix, we would like to write $\Gamma_{d,d}$ as a trace over two different basis. We first consider the case $d=1$,  before generalizing the result to arbitrary $d$.   

To evaluate 
\be
\Gamma_{1,1}=\Tr q^{{1\over 2}p_L^2}\bar q^{{1\over 2}p_R^2}\qquad \where \qquad p_{\overset{\scriptstyle L}{R}}={1\over \sqrt{2}}\left({m\over R}\pm Rn\right)\!,
\ee
the most commonly used basis of zero modes is $\abs m,n\rangle$, $m,n\in\Z$, which is orthonormal,
\be
\forall m,n, m', n'\, , \qquad  \langle m',n'\abs m,n\rangle = \delta_{m'm}\delta_{n'n}\, . 
\ee
Another basis can be described using the following definition : For any pair of integers $(m,n)\neq (0,0)$, if the first nonzero entry is positive, we say that  $(m,n)>0$, and otherwise we say that $(m,n)<0$. The new basis is then,
\be
\abs 0,0\rangle \, , \qquad \forall (m,n)>0\, , \; \forall \epsilon=\pm 1, \quad \abs m,n;\epsilon\rangle={\abs m,n\rangle+\epsilon\, \abs \!-m,-n\rangle\over \sqrt{2}}\, , 
\ee
which is also orthonormal : 
\begin{align}
&\langle 0,0\abs0,0\rangle=1\, ,\nonumber \\
&\forall (m,n)>0\, , \; \forall \epsilon=\pm 1, \quad\; \,\;\langle 0,0\abs m,n;\epsilon\rangle=0\, , \nonumber \\  
&\forall (m',n')>0\, , \; \forall \epsilon'=\pm 1, \quad \langle m',n';\epsilon'\abs m,n;\epsilon\rangle=\delta_{m'm}\delta_{n'n}\delta_{\epsilon\epsilon'}\, .
\end{align}
Both bases diagonalize the mass operator and level matching condition
\be
q^{{1\over 2}p_L^2}\bar q^{{1\over 2}p_R^2}=e^{2i\pi\tau_1mn}\, e^{-\pi\tau_2\left[({m\over R}^2+(nR)^2\right]}\, ,
\ee
so that
\be
\Gamma_{1,1}=1+\Gamma_{1,1}^++ \Gamma_{1,1}^-\, , \qquad \Gamma_{1,1}^+=\Gamma_{1,1}^-\, ,
\ee
where we have defined
\begin{align}
&\Gamma_{1,1}=\sum_{m,n} \langle m,n\abs q^{{1\over 2}p_L^2}\bar q^{{1\over 2}p_R^2}\abs m,n\rangle\, , \nonumber\\
&\Gamma_{1,1}^+=\sum_{(m,n)>0} \langle m,n;+1\abs q^{{1\over 2}p_L^2}\bar q^{{1\over 2}p_R^2}\abs m,n;+1\rangle\, , \nonumber\\
&\Gamma_{1,1}^-=\sum_{(m,n)>0} \langle m,n;-1\abs q^{{1\over 2}p_L^2}\bar q^{{1\over 2}p_R^2}\abs m,n;-1\rangle\, .
\end{align}
The use of the second basis is relevant to write the 1-loop partition function associated to an orbidold $S^1/\Z_2$, since the first basis does not diagonalize the $\Z_2$ twist generator $\boldsymbol{G}$, while the second one does,
\be
\forall m,n\, , \;\; \boldsymbol{G}\abs m,n\rangle=\abs \!-m,-n\rangle\, ; \qquad \forall (m,n)>0\, , \; \forall \epsilon=\pm 1\, , \;\; \boldsymbol{G}\abs m,n;\epsilon\rangle=\epsilon \abs m,n;\epsilon\rangle\, .
\ee
As a result, the two traces involved in the untwisted sector of the $S^1/\Z_2$ partition function can be written as, 
\be
\Tr \!\left[\boldsymbol{G}^G\, q^{{1\over 2}p_L^2}\bar q^{{1\over 2}p_R^2}\right]=1+\Gamma_{1,1}^++(-1)^G\, \Gamma_{1,1}^-\, , \qquad G=0,1 \mbox{ modulo 2}\, .
\ee
The first basis is useful for other purposes. The right-moving coordinate of the string in the direction of the circle of radius $R/\sqrt{\Ms}$ realizes in spacetime an $SU(2)$ gauge symmetry, whose charge operator is $Q\equiv p_R$. If the first basis diagonalizes $Q$, the second one does not : 
\begin{align}
&\forall m,n\, ,\qquad \quad\quad \qquad\qquad  \qquad Q\abs m,n\rangle={1\over \sqrt{2}}\left({m\over R}-nR\right)\!\abs m,n\rangle\, ,\nonumber \\
&\forall (m,n)>0\, , \; \forall \epsilon=\pm 1\, , \qquad Q\abs m,n;\epsilon\rangle={1\over \sqrt{2}}\left({m\over R}-nR\right)\!\abs m,n;-\epsilon\rangle\, .
\end{align}
For instance, as said in the text, the extra massless states arising  when $R=1$, namely $\abs m,n\rangle$, $m=-n=\pm 1$, have charges $Q=\pm \sqrt{2}$, which are the roots of an enhanced $SU(2)$ gauge symmetry, while the state $\abs 0,0\rangle$ uplifted by one level 1 oscillator of the circle coordinate provides the Cartan generator.  When $R\neq 1$, the only state remaining massless being the Cartan generator, $R$ parameterizes the Coulomb phase of the gauge theory, $SU(2)\to U(1)$. 

The $d$-dimensional case can be considered the same way. For arbitrary $2d$-tuple $(m_i,n_i)\neq (0,\dots,0)$, if the first nonzero entry is positive, we say that  $(m_i,n_i)>0$, and otherwise we say that $(m_i,n_i)<0$. The set of states
\be
\abs 0,\dots,0\rangle \, , \qquad \forall (m_i,n_i)>0\, , \; \forall \epsilon=\pm 1, \quad \abs m_i,n_i;\epsilon\rangle={\abs m_i,n_i\rangle+\epsilon\, \abs \!-m_i,-n_i\rangle\over \sqrt{2}}\, , 
\ee
form an orthonormal basis of zero modes, in term of which we have
\be
\Gamma_{d,d}=1+\Gamma_{d,d}^++ \Gamma_{d,d}^-\, , \qquad \Gamma_{d,d}^+= \Gamma_{d,d}^-\, , 
\ee
where we have defined 
\begin{align}
&\Gamma_{d,d}=\sum_{m_i,n_i} \langle m_i,n_i\abs q^{{1\over 2}\abs p_L\abs^2}\bar q^{{1\over 2}\abs p_R\abs^2}\abs m_i,n_i\rangle\, , \nonumber\\
&\Gamma_{d,d}^+=\sum_{(m_i,n_i)>0} \langle m_i,n_i;+1\abs q^{{1\over 2}\abs p_L\abs^2}\bar q^{{1\over 2}\abs p_R\abs^2}\abs m_i,n_i;+1\rangle\, , \nonumber\\
&\Gamma_{d,d}^-=\sum_{(m_i,n_i)>0} \langle m_i,n_i;-1\abs q^{{1\over 2}\abs p_L\abs^2}\bar q^{{1\over 2}\abs p_R\abs^2}\abs m_i,n_i;-1\rangle\,,
\end{align}
and $p_L,p_R$ are generalized momenta in $d$ dimensions (see Eq. (\ref{pLR}) for the case $d=2$, with $h=0$).  
Inserting the $\Z_2$ generator $\boldsymbol{G}$ in the traces, we have
\be
\Tr \!\left[\boldsymbol{G}^G\, q^{{1\over 2}p_L^2}\bar q^{{1\over 2}p_R^2}\right]=1+\Gamma_{d,d}^++(-1)^G\, \Gamma_{d,d}^-\, , \qquad G=0,1 \mbox{ modulo 2}\, ,
\ee
which justifies Eq. (\ref{z44}).



\end{document}